\documentclass[10pt,journal,compsoc]{IEEEtran}
\usepackage{textcomp}
\ifCLASSOPTIONcompsoc
  \usepackage[nocompress]{cite}
\else
  \usepackage{cite}
\fi
\ifCLASSINFOpdf
   \usepackage[pdftex]{graphicx}
\else
\fi
\usepackage{algorithmic}
\usepackage{algorithm}

\usepackage{array}

\ifCLASSOPTIONcompsoc
  \usepackage[caption=false,font=footnotesize,labelfont=sf,textfont=sf]{subfig}
\else
  \usepackage[caption=false,font=footnotesize]{subfig}
\fi
\begin{document}
%
\title{LSB: A Lightweight Scalable BlockChain for IoT Security and Privacy}
%
%
%
%

\author{Ali~Dorri, 
        Salil S.~Kanhere,
        Raja~Jurdak,
        and~Praveen~Gauravaram
\IEEEcompsocitemizethanks{\IEEEcompsocthanksitem Ali Dorri is with the School of computer science and engineering,  UNSW, Sydney, Australia. He is also with DATA61 CSIRO, Australia. 
\protect\\
E-mail: Ali.dorri@unsw.edu.au
\IEEEcompsocthanksitem Salil S. Kanhere is with the School of computer science and engineering, UNSW, Sydney, Australia. 
E-mail: Salil.kanhere@unsw.edu.au
\IEEEcompsocthanksitem Raja Jurdak is with DATA61 CSIRO, Brisbane, Australia.
E-mail: Raja.Jurdak@csiro.au
\IEEEcompsocthanksitem Praveen Gauravaram is with Tata Consultancy Services, Brisbane, Australia.
Email: p.gauravaram@tcs.com}
\thanks{}}

%
%

\markboth{}%
{Shell \MakeLowercase{\textit{et al.}}: Bare Demo of IEEEtran.cls for Computer Society Journals}
%



\IEEEtitleabstractindextext{%
\begin{abstract}
BlockChain (BC) has attracted tremendous attention due to its immutable nature and the associated security and privacy benefits. BC has the potential to overcome security and privacy challenges of Internet of Things (IoT). However, BC is computationally expensive, has limited scalability and incurs significant bandwidth overheads and delays which are not suited to the IoT context. We propose a tiered Lightweight Scalable BC (LSB) that is optimized for IoT requirements.  We explore LSB in a smart home setting as a representative example for broader IoT applications. Low resource devices in a smart home benefit from a centralized manager that establishes shared keys for communication and processes all incoming and outgoing requests.  LSB achieves decentralization by forming an overlay network where high resource devices jointly manage a public BC that ensures end-to-end privacy and security. The overlay is organized as distinct clusters to reduce overheads and the cluster heads are responsible for managing the public BC. LSB incorporates several optimizations which include algorithms for lightweight consensus, distributed trust and throughput management. Qualitative arguments demonstrate that LSB is resilient to several security attacks. Extensive simulations show that LSB decreases packet overhead and delay and increases BC scalability compared to relevant baselines.
\end{abstract}

}

\maketitle

\IEEEdisplaynontitleabstractindextext

%
\IEEEpeerreviewmaketitle

\IEEEraisesectionheading{\section{Introduction}\label{sec:introduction}}
 \IEEEPARstart{B}{lockchian} (BC) is an immutable timestamp  ledger of blocks that is used for storing and sharing  data in a distributed manner \cite{kosba2016hawk}. The stored data might be payment history, e.g. Bitcoin  \cite{nakamoto2008bitcoin}, or a contract \cite{wood2014ethereum} or even personal data \cite{yue2016healthcare}. In recent years, BC has attracted tremendous attention from practitioners and academics   in different disciplines (including law, finance, and computer science) due to its salient features which include distributed structure, immutability and security and privacy \cite{abramaowicz2016cryptocurrency}. A recent survey  \cite{BCInIndustry} has observed that BC is expected to impact at least 27 different industry sectors. \par 
BC maintains a distributed digital ledger of transactions that is shared across all participating nodes. New transactions are verified and confirmed by other nodes participating in the network, thus eliminating the need for a central authority.   Appending a new block  to the BC (referred to as mining in literature) entails solving a computationally demanding,  hard-to-solve, and easy-to-verify puzzle. This puzzle underpins a trustless consensus algorithm among untrusted nodes. The computation resources required to participate in the consensus algorithm can be very significant, which restricts the number of blocks that can be mined by a node and thus offers protection against malicious mining of blocks. Solving the puzzle involves a process which introduces randomness among nodes who wish to mine a new block (miners). Existing BC implementations typically use one of the following consensus algorithms:  Proof of Work (POW) \cite{vukolic2015quest} or Proof of Stake (POS) \cite{wood2014ethereum}. POW demands high computational resources, while POS demands both computational and memory resources to solve a cryptography puzzle. All communication between nodes  is encrypted which protects against eavesdropping. BC users employ changeable Public Keys (PK) that prevents them from being tracked, thus ensuring their privacy. \par    
 BC was first introduced in a cryptocurrency  known as Bitcoin  and since then has been widely used  in other cryptocurrencies known as altcoins \cite{Altcoin}. In recent years,  BC has attracted attention in non-monetary applications including but not limited to: sharing of healthcare data  \cite{yue2016healthcare}, securing robotic swarms \cite{ferrer2016blockchain} and verifying  proof of location \cite{brambilla2016using}. In this paper, we argue that  BC is an  effective technology for overcoming the  security and privacy challenges  that emerge from connecting billions of everyday devices to the Internet, i.e., the Internet of Things (IoT). Conventional security and privacy methods tend to be ineffective in IoT due to the following challenges:\par 
 \begin{itemize}
 \item Resource consumption: Most IoT devices have limited resources, including bandwidth, computation, and memory, which is incompatible with  the requirements of complex security  solutions  \cite{zhang2014iot}. 
 \item Centralization: Current IoT ecosystems rely on centralised brokered communication models where all devices are identified, authenticated and connected through cloud servers.  This model is unlikely to scale as billions of devices are connected. Moreover, cloud servers will remain a bottleneck and point of failure that can disrupt the entire network  \cite{zhang2014iot}.  
 \item  Lack of privacy: Conventional privacy preserving methods rely on  revealing noisy or summarized data  to the data requester \cite{de2014openpds}. In contrast, several IoT applications require users to  reveal precise data to the Service Providers (SP) to receive personalized services. 
 \end{itemize}
The various benefits afforded by BC technology as outlined earlier in this section make it an attractive solution for addressing the aforementioned problems in IoT. However, existing instantiations of BC cannot be readily adopted in the IoT context for the following reasons: \par 

\textit{ Complex consensus algorithms:}  The consensus algorithms employed in BC (POW or POS) require significant computational resources which are far beyond the capabilities of most IoT devices.\par 
 \textit{Scalability and overheads:} In a typical BC implementation, all blocks are broadcast to and verified by all nodes. This leads to significant scalability issues since the broadcast traffic and processing overheads would increase quadratically with the number of nodes in network. The associated overheads are intractable as most IoT devices have limited bandwidth connections (e.g. Low Power Wide Area Networks such as LoRa) and processing capabilities. \par 
\textit{ Latency: }There  is a non-trivial delay associated with ensuring that a transaction is confirmed by nodes participating in the BC. For example, in Bitcoin, it can take up to 30 minutes for a transaction to be confirmed.  Most IoT applications have stricter delay requirements e.g.  a service provider requesting data from a smart home sensor should not have to wait for several minutes.   \par 
\textit{ Security overheads:} Some of the compute-intensive security mechanisms in conventional BCs provide protection against double spending, which are appropriate for cryptocurrency but not directly applicable in the IoT context. \par 
\textit{ Throughput:} In BC, the throughput is defined as the number of transactions that can be stored. Classical instantiations of BC have limited throughput. For example, Bitcoin throughput is 7 transactions per second. However, the number of transactions in the IoT ecosystem would far exceed such limits due to extensive interactions between various entities. \par 
In this paper, we propose a Lightweight Scalable BC (LSB) for IoT security and privacy  that addresses the above  issues.  We use a smart home setting for illustrative purposes but LSB is application agnostic and well-suited for diverse IoT applications. The framework consists of two main tiers namely, smart home and overlay. A transaction is defined as the basic communication primitive for exchanging information among any entities. To optimize resource consumption, IoT devices in the local smart home utilize   a local private Immutable Ledger (IL) of local transactions, that is structurally similar to BC  but is managed centrally. Symmetric encryption is used to encrypt transactions in this tier.  The overlay tier comprises of capable nodes, such as  SP servers, that collaboratively manage a public BC which stores overlay transactions. To ensure scalability, the overlay nodes are organized as clusters and only the Cluster Heads (CH) are responsible for managing the public BC. We propose a lightweight consensus algorithm that limits the number of new blocks generated by the CHs within a tunable consensus period. \par 
To reduce the computation overhead associated with verifying new blocks that are to be added to the public BC, LSB employs a distributed trust algorithm. Each CH accumulates evidence about other CHs based on the validity of new blocks that they generate. The number of transactions in a new block that need to be verified is gradually reduced as CHs accrue  trust in each other. Finally, we propose a Distributed Throughput Management (DTM) mechanism to dynamically adjust certain system parameters to ensure that the throughput of the public BC (i.e., the number of transactions appended to the BC) does not significantly deviate from the transaction load in the network. DTM ensures that network is self-scaling, i.e., as the network grows in size, more transactions  can be appended to the public  BC, thus increasing the throughput. In LBS, the flow of data to and from IoT devices is kept separate from the transaction flow. Transactions are broadcast among the overlay nodes while data packets are routed toward their destination. This separation allows optimal unicast routing of data packets, thus resulting in reduced delays. \par 
The key contributions of this paper are summarized below:
\begin{enumerate}

\item We present a comprehensive tiered framework based on BC technology for preserving security and privacy for IoT. We use a smart home setting as the basis for presenting the work. However, our ideas are application agonistic and are well suited to a broad range of IoT applications.
\item  Our instantiation of the BC, known as Lightweight Scalable Blockchain (LSB), is tailored to meet the specific requirements of IoT devices and applications. We incorporate a number of optimizations which include a lightweight consensus algorithm, a distributed trust method, a distributed throughput management strategy and a separation of the transaction traffic from the data flow.
\item We undertake a qualitative analysis of LSB against 12 relevant cyber attacks and outline the specific defence mechanisms, which ensure that LSB is resilient against all of them. Additionally, a risk analysis is conducted to investigate the likelihood of the attacks.
\item  We conduct extensive simulations using Cooja and NS3 to evaluate key performance parameters including latency, processing time, and resilience against cyber attacks. Our results justify various design decisions made and demonstrate the efficacy of the proposed optimizations.
\end{enumerate}
This paper is a significant extension of the preliminary ideas presented in our prior works \cite{dorri2017towardsBC,dorri2017BC} including new concepts, algorithms and extensive evaluations.\par 
The rest of the paper is organized as follows. The proposed architecture and details of two tiers are discussed in Section  \ref{sec:def&cons}. An overview of transactions  is discussed   in Section \ref{sec:transaction-flow}. Detailed security analysis and performance evaluations are presented in  Section \ref{sec:Evaluations}. Section \ref{discussion} discusses further aspects of LSB.  Section \ref{relatedwork} presents a  literature review on IoT security and BC applications, and  finally Section \ref{conclusion} concludes the paper and outlines and future work. 
\section{Lightweight Scalable Blockchain (LSB)}\label{sec:def&cons}
In this section, we  discuss the overlay and smart home tiers in detail. We begin by defining two fundamental concepts:  
\begin{itemize}
\item \textit{Transaction:} The basic communication primitive for exchanging control information among any entities is referred to as a transaction. As noted in Section \ref{sec:introduction}, the data flow is distinct from transactions. 
\item \textit{BlockManager (BM): } BM is an entity that is responsible for managing the BC. This includes generation, verification and storage of individual transactions and blocks of transactions. BMs in the overlay and smart home tiers have slightly different functions as will be explained in the subsequent sections.  
\end{itemize}
\subsection{Overlay}\label{sec:sub-overlay}
\begin{figure*}[h]   	
\begin{center}
\includegraphics[width=15cm ,height=8cm ,keepaspectratio]{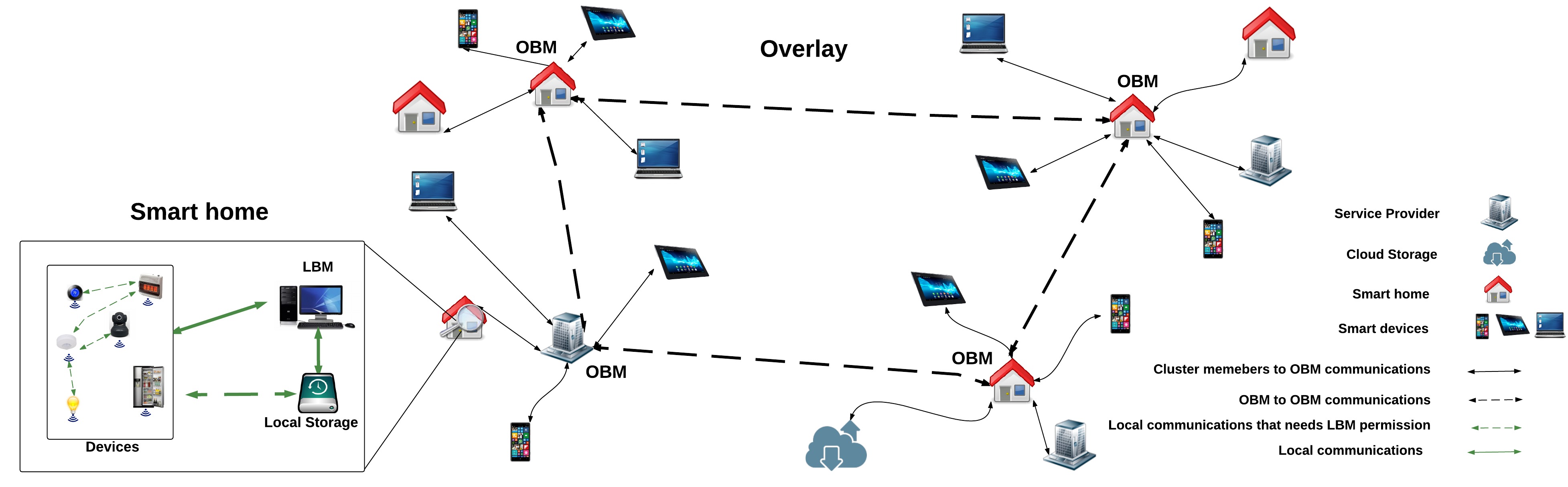}   
\caption{An overview of LSB.}
\label{fig:overal-pic}
\end{center}
\end{figure*}

Similar to Bitcoin, we assume that each node in the overlay is known by a   Public Key (PK). Nodes use a fresh PK  to generate   each new transaction to ensure anonymity (discussed further in Section \ref{sec:sub:security-analysis}).  The overlay, as shown in Fig  \ref{fig:overal-pic},  is comprised of various entities, known as overlay nodes, including the smart home (represented by the Local BM (LBM), which will be introduced in Section \ref{sec-smarthome}), mobile devices, Service Provider (SP) servers, and cloud storage (used by smart home devices for storing data). \par 
The overlay network could potentially consist of a large number of nodes. Thus to ensure scalability, we assume that the public BC is managed by a subset of the overlay nodes. We assume that a clustering algorithm such as in \cite{kousaridas2015systas} is used to group nodes into clusters, with each cluster electing a Cluster Head (CH).  CHs are responsible for managing the BC and are thus referred to as Overlay Block Mangers (OBMs). Additionally, CHs process incoming and outgoing transactions that are generated to or from their cluster members.    A node selected as a CH  is expected to  remain online for an extended duration of time and to have sufficient resources for processing blocks and transactions.  We assume that mechanisms such as those used in \cite{stoica2001chord} for managing CH failures are in place.  Since the fundamental tasks are performed by the CHs, LSB is unaffected by IoT device dynamics, i.e., joining and leaving of devices. \par 

Transactions generated by an overlay node are secured using asymmetric encryption, digital signatures and cryptographic hash functions (e.g. SHA256). Transactions in the overlay can be further classified into (i) \textit{single signature} transactions which only contain the signature of the transaction generator and (ii) \textit{multisig} transactions which are signed by both the transaction generator (requester) and  receiver (requestee). The majority of transactions in LSB are multisig. The instance where single signature transactions are used is discussed in Sections \ref{sec:transaction-flow} and \ref{discussion}.\par
\begin{figure}[h]   	
\begin{center}
 \subfloat[]{\label{fig:overlay-trans-struc} \includegraphics[width=5cm,height=5cm,keepaspectratio]{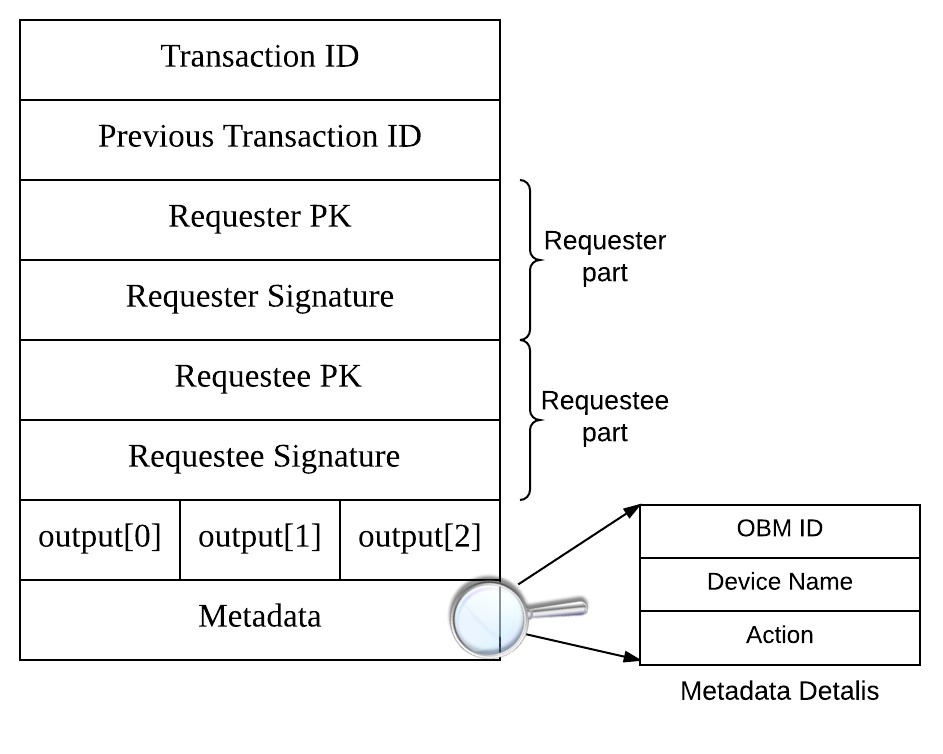} }\hfill
 \subfloat[]{\label{fig:trust-overlay-pic} \includegraphics[width=5cm,height=5cm,keepaspectratio]{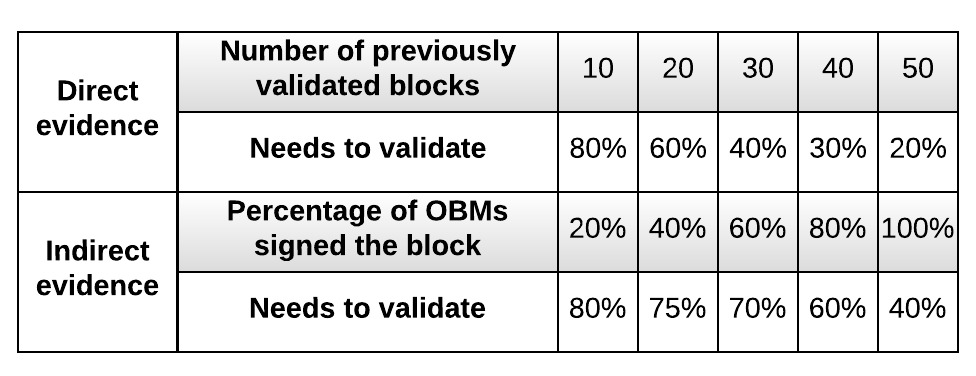} }\\
\caption{  a) Structure of the multisig transactions, b) Trust table.}
\end{center}
\end{figure}

 
The structure of a multisig transaction is shown in Fig \ref{fig:overlay-trans-struc}. The first field is an identifier for the transaction while the second field is a pointer to the previous transaction of the same requester node. Thus, all transactions created by a requester  are chained together. This is followed by the PK and signature of the requester and requestee. The latter signature is appended when  the transaction is received by the requestee. The seventh field is the transaction output and is set by the requester. The output field contains the following 3 entries: (i) the total number of transactions generated by the requester that have been accepted by the requestee (ii) the total number of transactions rejected by the requestee (iii) the hash of the PK that the requester will use for its next transaction. The first two fields provide historical information that is necessary for computing the reputation of the requester, which is used in the distributed trust algorithm outlined in Section \ref{sec:sub:verification}. The last output field is  necessary for future verification of the requester, since the overlay nodes change the PK used for generating each new transaction.  The final field in a multisig transaction, i.e. "metadata", provides information about the desired action (a detailed discussion is in Section  \ref{sec:transaction-flow}) and the smart home device which is the target of this action.  A single signature transaction has a similar structure but excludes the requestee PK and signature,  metadata, and outputs [0] and [1] as only one  overlay node is involved. Note that, multisig and single signature transactions are organized as  separate ledgers since the respective outputs are different. \par 

The key transactions in the overlay are as follows: \par 
\begin{itemize}
\item Genesis:  Each overlay node must first create a genesis transaction which serves as the starting point for its ledger in the public BC. The genesis transaction generation is discussed in section \ref{sec:transaction-flow}.  
\item Store: An overlay node generates a store transaction to store data in the cloud storage.  
\item Access: An access transaction is generated by an overlay node to request  stored data of a device, e.g., a SP (requester) may request all data stored by a device (requestee) for the past week. 
\item Monitor: A  monitor transaction is generated by an overlay node, e.g., a SP may wish to obtain real-time data from a device.
\end{itemize}
In LBS, the data flow is kept separate from the transaction flow. Thus, in response to an access or monitor transaction, the requestee device sends the data to the requester in a separate data packet(s) after it is confirmed that the requester is authorised to access the data. Similarly, for the store transaction the data created by the requester is sent distinctly from the transaction. Unlike transactions which are broadcast, the data packets are unicast and can be routed along optimal paths through the overlay network using routing protocols such as OSPF  \cite{OSPFref}.\par 
The overlay transactions are stored in the the public BC that is managed by the OBMs. Each block in the BC consists of two main parts namely, transactions  and block header. The block header contains the following:  hash of the previous block, block generator ID, and signatures of the verifiers. The hash of the previous block in the public BC  ensures immutability. If an attacker attempts to change a previously stored transaction, then the hash of the corresponding block which is stored in the next block will no longer be consistent and will thus expose this attack.  The "block generator ID" and "signatures of the verifiers" fields  will be discussed later in this section. Similar to Bitcoin, multiple transactions are  grouped together and then processed as one block. A block can store at most $T\_max$ transactions. The value of $T\_max$ affects the BC throughput in a way that with a  larger $T\_max$, more transactions can be stored in a single block.  \par 
When an  OBM receives transaction Y, it first checks whether the requestee of this transaction is contained within its cluster. An OBM maintains a key list (essentially a simplified access control list) consisting of requester/requestee PK pairs which indicates the requesters that are allowed to send transactions to specific requestees.  This key list  is updated by a cluster member to give permission to other overlay nodes to send transactions to it. A requestee may set the requester value in the OBM key list as "broadcast", which implies that it will receive all transactions that contain its PK as the requestee PK. If the requester and requestee of the incoming transaction Y match with an entry in the key list, then  the OBM sends the transaction to the requestee (which is within its cluster and thus directly connected to the OBM). If the requestee in Y does not belong to the OBM's cluster, then the transaction is broadcast to all other OBMs. All pending transactions are stored in a transaction pool at each OBM. When the size of the running pool becomes equal to $T\_max$ the OBM starts the process of creating a new block  using a consensus algorithm.\par 
\subsubsection{Consensus algorithm}\label{sec:consensus}
As noted in Section \ref{sec:introduction}, in LSB, we propose a time-based consensus algorithm in place of the more resource-intensive alternatives such as PoW and PoS that are typically used in BC. The consensus algorithm must ensure a block generator is selected randomly among nodes and is limited in the number of blocks it can generate. To introduce randomness among block generators, each OBM must wait for a random time, known as \textit{waiting-period}, prior to generating a new block. Since the waiting-period differs for each OBM, an OBM might receive a new block created by another OBM that contains some or all of the  transactions that are currently within the pool of transactions of the OBM. In this instance, this OBM  must remove these transactions from its pool as they are stored in the BC by another OBM. Requiring OBMs to wait for a random time also reduces the number of duplicate blocks that can be generated simultaneously. The maximum waiting-time is capped at twice the maximum end-to-end delay in the overlay network. When a new block is generated, it is broadcast to all other overlay nodes so that it can be appended to the BC.\par
To protect the overlay against a malicious OBM that may potentially  generate a large number of blocks with fake transactions leading to an appending attack (discussed in Section \ref{sec:sub:security-analysis}), the periodicity  with which an OBM can generate blocks is restricted such that  only one block can be generated over an interval denoted by \textit{consensus-period}. The consensus-period is adjusted by Distributed Throughput Management (DTM) and is discussed in  Section \ref{sub-sec:dis-throughput}. The default (and maximum) value for consensus-period is 10 minutes, which is similar to the mining interval in Bitcoin.  The  minimum value of consensus-period is equal to twice the maximum end-to-end delay in the overlay, to ensure that there is sufficient time for disseminating a new block generated by other OBMs. Each OBM monitors the frequency with which other OBMs generate blocks. Any non-compliant blocks are dropped and the trust associated with the responsible OBM is decreased as outlined in the following sub-section. To prevent OBMs from always claiming to have a short waiting-period, the neighbor OBMs monitor the frequently that an OBM generates new blocks in the begening of the waiting-period. If the number of such blocks exceeds a threshold, defined based on application by the BC designers, the OBMs drop the block generated by their neighbor.  \par 

\subsubsection{Verification}\label{sec:sub:verification}
An OBM must validate each new block that it receives from other OBMs prior to appending it to the BC. To validate a block, the OBM first validates the signature of the block generator. It is assumed that each OBM uses a  pre-defined key for generating blocks and it is assumed that these keys are known to all other OBMs \cite{lu2008framework}. Next, each individual transaction in the block is verified. A block is considered to be valid if all transactions contained in the block are valid.   \par 
Algorithm \ref{algo-verification} outlines the procedure for verifying an individual transaction, X. Recall that in the public BC all multisig transactions generated by each requester are  organized in a separate ledger. The output of the multisig transactions creates a   reputation metric for the requester. As discussed in Section \ref{sec:sub-overlay}, the link between successive transactions is established by including the hash of the PK that will  be used by the requester for the next transaction in the third output field of the current transaction. Thus, the OBM first confirms this by comparing  the hash of the requester PK in X with  output[2] of the previous transaction of this requester (lines 1,2). Following this, the requester signature, that is contained within the fourth field of X,   is verified (also called redeemed) using its PK in X (lines 4,5). Recall from Section \ref{sec:sub-overlay} that output [0] and output [1] are used to create reputation for the requester node. Initially, the requester sets these outputs (based on its history of transactions) in the multisig transaction. If the requestee accepts the transaction, then it would increase the output [0] by one. Otherwise, the requestee increases the output [1]. To protect the BC against nodes that claim false reputation by increasing their outputs before sending them to the requestee, in the next step of transaction verification, the OBM checks that only one of  X's outputs, i.e. either the number of successful transactions (i.e. output [0]) or the number of rejected transactions (i.e. output [1]), is increased  only by one (lines 8,9).  Following this, the requestee signature  is verified using its PK in X (lines 11,12). If the steps complete successfully, X is verified. \par 
\begin{algorithm}[h]
 \caption{Transaction verification.}
 \begin{algorithmic}[1]
 \algsetup{linenosize=\tiny}
  \scriptsize
 \renewcommand{\algorithmicrequire}{\textbf{Input:} }
 \renewcommand{\algorithmicensure}{\textbf{Output:}  }
 \REQUIRE Overlay Transaction (X)
 \ENSURE  True or False
  \\ \textit{Requester verification} :
    \IF {(hash (X.Requester-PK)  $\ne  $ $X_{-1}$.output[2])}
  \RETURN False; 
    \ELSE 
    	\IF {(X.requester-PK \textit{redeem} x.requester-signature)}
    	\RETURN False; 
    	\ENDIF
  \ENDIF
     \\ \textit{Ouput validation} :     
   \IF {(X.output[0] -  $X_{-1}$.output[0]) + (X.output[1] -  $X_{-1}$.output[1])$ >$ 1)}
   \RETURN False; 
    \ENDIF
     \\ \textit{Requestee verification} :
    \IF {(X.requestee-PK \textit{redeem} x.requestee-signature)}
  \RETURN true; 
    \ENDIF
 \end{algorithmic}
 \label{algo-verification}
 \end{algorithm}
 Verifying all transactions and blocks is computationally demanding, particularly when the number of nodes in the overlay network increases. In the IoT context, one can expect serious scalability issues since the number of nodes is expected to be very large. To address this, LSB uses a \textit{distributed trust algorithm} that gradually reduces the number of transactions that need to be verified in each new block as OBMs build up trust in one another. The algorithm introduces the notions of direct and indirect evidence as follows:\par 
\textit{Direct evidence: } OBM \textit{A} has direct evidence about  OBM \textit{B} if it previously verified at least one block that was generated by \textit{B}. \par 
\textit{Indirect evidence: } If OBM \textit{A} does not have direct evidence about OBM \textit{B}, but if one of the other OBMs has confirmed that the block generated by \textit{B} is valid, then \textit{A} has indirect evidence about \textit{B}.\par 
Each OBM maintains a list that records pertinent information to establish direct evidence. For this, the OBM records the number of blocks it has validated for every other OBM. Recall from Section \ref{sec:consensus} that an OBM might create blocks which are non-compliant with the consensus algorithm. Other OBMs that receive a non-compliant block will drop the same and decrement the direct trust associated with the responsible OBM by one. If the malicious OBM continues with this behaviour, its trust rating would be correspondingly reduced. This implies that more and more of its transactions will have to be verified by the other OBMs.  For indirect evidence, the OBM checks the number of other OBMs that have verified a received block generated by an OBM. The core idea behind the distributed trust algorithm is that the stronger the evidence an OBM has gathered about the OBM generating the new block, fewer transactions within that block need to be verified to validate the block. A trust table, an example of which is illustrated in Fig \ref{fig:trust-overlay-pic} is maintained by the OBMs to implement this strategy. Direct evidence takes precedence over indirect evidence. If the OBM has direct evidence about the block creator, then a fraction of the transactions within the block are selected to be validated as per Fig \ref{fig:trust-overlay-pic}. In the case that there is no direct evidence, the OBM checks if indirect evidence is available and then selects a different fraction of transactions based on how many other OBMs have vouched for the block generator as per Fig \ref{fig:trust-overlay-pic}. Note that, a certain fraction of transactions are always verified even if there is strong evidence to protect against a potentially compromised OBM. If no evidence is recorded, then all transactions in the block are verified. \par 
\subsubsection{Distributed Throughput Management (DTM)} \label{sub-sec:dis-throughput}
The classical consensus algorithms used in BC limit the BC throughput, which is measured as the number of transactions stored in the BC per second, as solving the cryptographic puzzle is computationally demanding. For instance, Bitcoin BC is limited to 7 transactions per second because of POW \cite{nakamoto2008bitcoin}. For IoT, such limits would be unacceptable, since there would be numerous interactions (and thus transactions) between various nodes. In LSB we propose a Distributed Throughput Management (DTM) mechanism (outlined in Algorithm \ref{algo-DTM} below) to actively monitor the BC utilization and make appropriate adjustments to ensure that it remains within an acceptable range. At the end of every consensus-period, each OBM computes the utilization ($\alpha$) as the ratio of the total number of new transactions generated to the total number of transactions added to the BC. Note that, since all transactions and blocks are broadcast to all OBMs, the utilization computed by all OBMs should be similar. The aim of DTM is to ensure that $\alpha$ remains within a certain desirable range ($\alpha_{min}$, $\alpha_{max}$). \par 
 \begin{algorithm}[h]
 \caption{Distributed Throughput Management.}
 \begin{algorithmic}[1]
 \algsetup{linenosize=\tiny}
  \scriptsize
 \renewcommand{\algorithmicrequire}{\textbf{Input:} }
 \REQUIRE  $\alpha $
 \WHILE {true}
    \IF {( $\alpha  > \alpha_{max} $)}
        \item  compute consensus-period$_{new}$  from Equation \ref{formula} with  $\alpha $ = $\frac{\alpha_{min} + \alpha_{max}}{2}$ 
 		\IF {(consensus-period$_{min}$ $<=$ consensus-period$_{new}$  ) }
 		 	\item update consensus-period to consensus-period$_{new}$
        \ELSE 
    	\item reset consensus-period to default value
    	\item compute $M$ from Equation \ref{formula} with  $\alpha $ = $\frac{\alpha_{min} + \alpha_{max}}{2}$ 
    	\item recluster overlay
    	\ENDIF
  \ENDIF
    	 \IF {( $  \alpha  < \alpha_{min}$)}
        \item compute consensus-period$_{new}$  from Equation \ref{formula} with  $\alpha $ = $\frac{\alpha_{min} + \alpha_{max}}{2}$ 
 		\IF {( consensus-period$_{new}$  $ <=$ consensus-period$_{max}$) }
 		 	\item update consensus-period to consensus-period$_{new}$
        \ELSE 
    	\item reset consensus-period to default value
    	\item compute $M$ from Equation \ref{formula} with  $\alpha $ = $\frac{\alpha_{min} + \alpha_{max}}{2}$ 
    	\item recluster overlay
	   \ENDIF
  \ENDIF    
    \ENDWHILE
 \end{algorithmic}
 \label{algo-DTM}
 \end{algorithm}
Assuming a network with $N$ nodes of which $M$ are OBMs and $R$ representing the average rate at which a node generates new transactions per second ($R$ can be estimated from the total number of transactions generated in the consensus-period), the utilization can be represented as follows:  
\begin{equation}\label{formula}
  \alpha = \frac{ N * R * Consensus-period  }{ T\_max * M}  
\end{equation}
The above equation suggests that there are two ways by which the utilization can be adjusted: (i) changing the consensus-period, which dictates the frequency with which blocks are appended to the BC; or (ii) changing $M$, as each OBM can generate one block within the consensus-period. The latter approach incurs significantly greater overheads as it requires reconfiguration of the entire overlay network (see Section \ref{sec:sub-overlay}). Thus, if $\alpha$ exceeds $\alpha_{max}$, in the first instance, DTM checks whether the consensus-period can be reduced. If so, then the new value for the consensus-period is computed using Equation \ref{formula} and assuming that $\alpha$ is equal to the mid-point of the desired range ($\alpha_{min}$, $\alpha_{max}$), which ensures a stable operating point for the network (line 2-3, Algorithm 2). On the contrary, if the consensus-period cannot be reduced then the network needs to be  reclustered  with a new value for $M$ (line 4). This new value is computed using Equation \ref{formula}, with $\alpha$ again set to the mid-point of the desired range and the consensus-period set to the default value, which is \textit{consensus-period$_{max}$}. This feature allows LSB to scale well, where an increased number of participating nodes delivers higher throughput.  We reset the consensus-period to default value  as it would otherwise remain unchanged at the minimum threshold and thus always require network reconfiguration if the utilization increased above its threshold. \par 
In the instance when the utilization drops below $\alpha_{min}$ an inverse approach is adopted, i.e. DTM first attempts to increase the consensus-period, otherwise it decreases the number of OBMs (lines 7 - 9, Algorithm \ref{algo-DTM}). \par
To ensure that all nodes are consistent about the action to be taken (whether it be changing the consensus-period or $M$), each OBM waits for a random duration and broadcasts a message specifying the action to be taken to all other OBMs. A recipient OBM checks whether the action is consistent with its decision. If so, it signs the original message and broadcasts it to other OBMs. If not, then it creates a fresh message specifying its action and broadcasts it to other OBMs. The action message that receives signatures from more than half the number of OBMs is assumed to be agreed-upon decision which all OBMs must follow. Note that, in most instances the actions taken by all nodes will be consistent. However, occasionally  there may be slight discrepancies in the OBMs estimate of the number of generated transactions due to packet loss or latency issues, which may in turn lead to minor differences in the computed consensus-period or $M$. In the rare event that there is no clear majority, the OBMs employ an election method such as in \cite{ongaro2014search} to reach a final agreement about the new consensus-period. \par 
In the event where the number of OBMs are to be changed, the network is reclustered  using the same clustering method used initially as discussed in Section \ref{sec:sub-overlay}.

\subsection{Smart home} \label{sec-smarthome}
The smart home is comprised of a variety of IoT devices which are managed by a Local BM (LBM).  Since IoT devices are typically resource-constrained,  local transactions  are encrypted using symmetric encryption, for which a shared key is established between the two parties,  and use lightweight cryptographic hash function, such as in \cite{Bogdanov2011}. In each smart home, the LBM centrally manages the local Immutable Ledger (IL) which is similar in structure to a BC,  and processes  local transactions and overlay transactions that are generated to or from the smart home.  The LBM  could be integrated with the  Internet gateway or a stand-alone middlebox such as F-secure \cite{fsecuresense} which acts as an intermediary between the IoT devices and the gateway.   The LBM uses the  generalized Diffie-Hellman \cite{delfs2002introduction} key distribution method to generate and distribute a shared key between two  local entities that are permitted to share data based on the policy header in local IL which is discussed later in this section.\par
\begin{figure*}[h]   	
\begin{center}
\includegraphics[width=15cm,height=8cm,keepaspectratio]{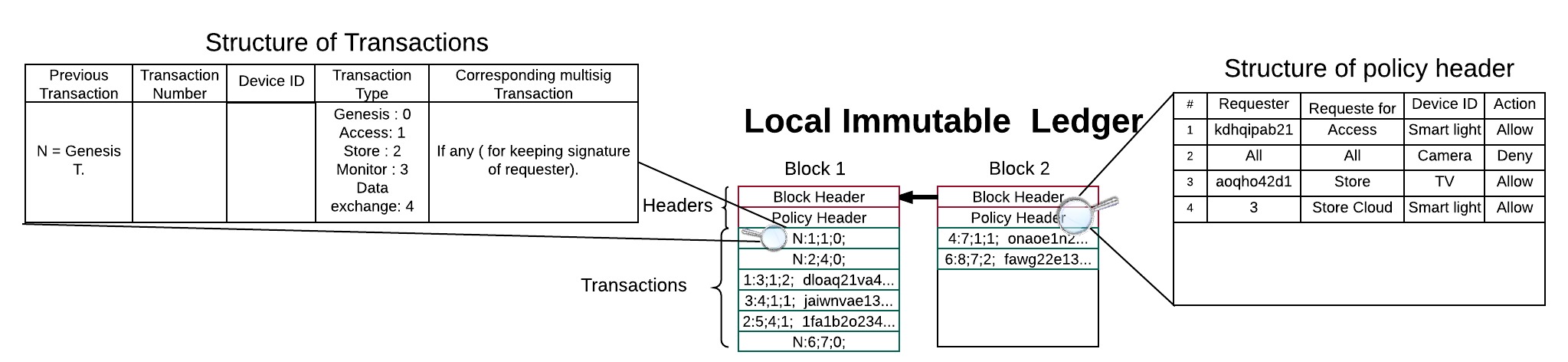}
\caption{The structure of the local Immutable Ledger (IL).}
\label{fig:local-bc}
\end{center}
\end{figure*}

 
The local IL records all local transactions and the overlay transactions for which the requestee is the LBM. As shown in Fig \ref{fig:local-bc}, each block in the local IL contains a block header and a policy header. The block header maintains the hash of the previous block to ensure  immutability similar to the public BC as discussed in Section \ref{sec:sub-overlay}.  The policy header is in the form of an Access Control List (ACL), which defines rules for processing local and overlay transactions. As shown in the right corner of Fig \ref{fig:local-bc}, the policy header has four parameters. The "Requester"  refers to the ID of the entity that generates the transaction. For incoming overlay transactions which are mutisig transactions, this should be the PK of the requester. For local transactions this refers to the specific IoT device (see Fig \ref{fig:local-bc} for an example). The second parameter in the policy header, indicates the permitted action (contained in the metadata of the transaction), which can be one of the following: \textit{store locally}, \textit{store cloud}, \textit{access}, \textit{monitor}, and \textit{monitor periodic}. The third parameter specifies  the target device. \par 
 
All  transactions are stored in the transaction part of IL for auditing. For each transaction, five fields are stored  as shown in the left side of Fig \ref{fig:local-bc}. The first  field is a pointer to the previous transaction of the same device which creates a chain of transactions for that specific device used by LBM for auditing and authentication. The second field is the transaction  identifier. The third field is the device ID.  For local transactions, this is the ID of the device that generates the transaction. For the overlay transactions, this is the ID of the device whose data is requested (by the requestee)  and is extracted from the metadata field of the received overlay transaction. "Transaction type" refers to the type of transaction (will be discussed in Section \ref{sec:transaction-flow}). If the transaction is sent by an overlay node, then the hash of the transaction, i.e. the transaction ID, is stored in the fifth field so that it can be used to refer to the transaction in the public BC. \par 
Each smart home  is equipped with  a local  storage repository which  is a  device such as a backup drive that is used by smart home devices to store  data locally. This storage can  be integrated with the LBM or it can be a separate device.  It is assumed that the  local storage is secure.  Assuming that a smart device is allowed to store data to the local storage (which is verified by checking the policy header), the LBM generates a shared key that is used by the device to authenticate with the storage.\par 
\section{Overview of Transactions} \label{sec:transaction-flow}
Having discussed the details of each tier, in this section we discuss the interactions between these tiers which is facilitated by different types of transactions. \par
\subsection{Local transactions}
Transactions exchanged between entities that belong to the same smart home are referred to as local transactions. Local transactions are encrypted by a shared key between the two entities involved in the transaction. The LBM generates shared keys for devices, using the generalized Diffe Hellman protocol \cite{delfs2002introduction}, after  seeking approval of the home owner.    To deny permission, the LBM  marks the previously distributed key as invalid by sending a control message to those entities that use  the key for their transactions. Local transactions are: \par 
\textit{Genesis transaction:} Each device first requires a genesis transaction in the local IL.  The LBM generates a shared key to encrypt communications between the device and the LBM and stores this key in the genesis transaction. \par 
\textit{Store locally:} Initially, the device that demands  to store data locally sends its request to the LBM. The LBM generates and distributes a shared key between the device and the local storage.  Local storage  uses the shared key for authentication. For further communications, the device and the storage communicate directly using the shared key.  \par 
\textit{Data exchange:} Each device inside the smart home may request data from another internal device to offer certain services e.g. the light sensor requests motion sensor data to turn the light on when someone enters home. To achieve the home owner control over local communications, first a shared key should be allocated  by the LBM  to the devices that request to share data.  After receiving the key, devices  communicate directly as long as the key is valid.\par 
 \subsection{Overlay transactions}
Transactions in which the transaction generator (requester) and the transaction receiver (requestee) are both overlay nodes are referred to as overlay transactions.   Recall that  asymmetric encryption is used for overlay transactions. The key overlay transactions are as follows. \par 
\textit{Genesis transaction:} Each overlay node must first create a genesis transaction in the public BC.  In the overlay tier, the genesis transaction is initiated by the overlay node using one of the following approaches:
\begin{itemize}
\item Certificate  authorities: In this approach, the node relies on the widely deployed Public Key Infrastructure (PKI) \cite{cooper2008internet} in the Internet. The overlay node contacts a trusted Certificate Authority (CA), which ratifies the node's PK by attaching a signed certificate. The node includes the certificate in the genesis transaction. To verify the transaction, an OBM verifies the certificate. It is assumed that the OBMs have access to a list of trusted CA root certificates for verification (similar to browsers and OSes). 
\item Burn coin in Bitcoin: Alternatively, if the node does not wish to rely on PKI, then it can privately create a genesis transaction by burning Bitcoins \cite{tschorsch2015bitcoin}. The node creates a permanent transaction in the Bitcoin BC by destroying a specified amount of coins (that can be defined as a design choice), which is referred to as "burning coins". The address of the burn transaction is used as the input of the genesis transaction.  The overlay node creates a genesis transaction with the same PK as the  burn transaction and sends it to the OBM whose cluster it belongs to. If the genesis transaction generator is an OBM, then it broadcast the transaction to other OBMs. To verify the received genesis transaction, the OBM matches the PK of the genesis transaction with the PK of the burn transaction in the Bitcoin BC. Next, the OBM verifies the  signature in the genesis transaction. 
\end{itemize}
In both approaches after verification, the OBM broadcast the genesis transaction  to other OBMs to be stored in the public BC. \par
\begin{figure*}[h]   	
\begin{center}
\includegraphics[width=15cm ,height=10cm ,keepaspectratio]{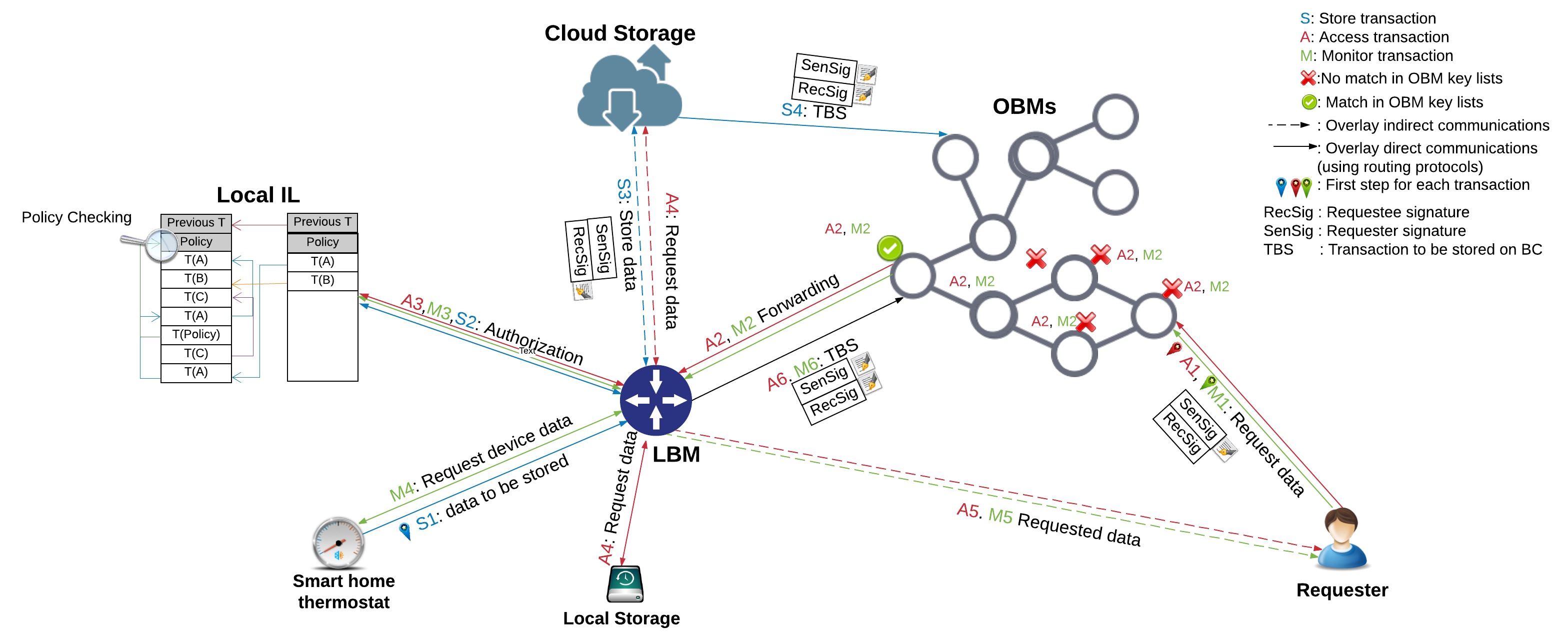}   
\caption{The process of store, access, and monitor transactions.}
\label{fig:transactions-flowchart}
\end{center}
\end{figure*}

 
\textit{Store cloud: }  We assume that a user who wishes to store his data, e.g. smart thermostat data, in the cloud has created an account with a cloud storage provider (e.g., Dropbox, OneDrive, etc.) out-of-band (i.e. independent of LSB). We assume that the user creates a public/private key pair for this cloud storage account and that the corresponding public key is used in  subsequent store cloud and access transactions. Recall that LSB creates a clear distinction between the control plane and the data plane to ensure that  the data packets can be routed efficiently through the network. To facilitate this, we assume that the LBM of the user sends a request during the initial setup to the cloud storage with the aforementioned PK. Upon authentication, the cloud storage sends the ID of its OBM to this LBM. Subsequently, all data being stored in the cloud can be directly routed to the cloud (as discussed in Section 2.1). The flow of events for the store cloud transaction are shown in Fig \ref{fig:transactions-flowchart}. The device that wishes to store data in the cloud sends a store cloud transaction to the LBM (S1 in Fig \ref{fig:transactions-flowchart}).  After authorization (S2)  the LBM sends data with the OBM ID of the cloud to its own OBM. The OBM then routes data directly to the cloud storage using  routing protocol.  After storing data, the cloud storage signs the received transaction from the LBM and sends it to its own OBM to be stored in the BC (S4).  \par 
\textit{Access and monitor:}  The flow of access and monitor transactions are shown in Fig \ref{fig:transactions-flowchart}. For both transactions,  the requester generates and sends the transaction to its OBM (A1, M1 in Fig \ref{fig:transactions-flowchart}). The OBM checks the keylists to find a match and if not then broadcast the transaction to other OBMs (A2, M2). By finding the match, the OBM forward the transaction to the LBM (A3, M3). The fourth step differs for access or monitor transaction. For access transaction, the data is fetched from either the local or cloud storage (A4), while for monitor transaction the real time data is requested from the device (M4). After receiving data,  the LBM routes it to the requester (A5, M5). Recall that the data flow is routed directly and separate from the transaction flow. Finally, the LBM signs the received transaction from the requester and sends it to its OBM to be stored in the public BC (A6, M6). 

\section{Evaluation and Discussion}\label{sec:Evaluations}
In this section we provide qualitative security and privacy analyses as well as quantitative performance evaluation. 
\subsection{Security and privacy analysis}\label{sec:sub:security-analysis}
In this section, we discuss LSB security,  privacy, and fault tolerance.  It is assumed that the adversary (or cooperative adversaries) can be the OBM, a device in the smart home, a node in the overlay network, or the cloud storage. Adversaries are able to sniff communications, discard transactions, create false transactions and blocks, change or delete data in storage, analyze multiple transactions in an attempt to deanonymize a node, and sign fake transactions to legitimize colluding nodes. We assume that standard secure encryption methods are used in the smart home and overlay tiers, which cannot be compromised by adversaries.   \par 

\textit{\textbf{Security:}} Table \ref{tab:security-req} summarizes the various mechanisms that allow LSB to meet key security requirements.   \par 
\begin{table*}[h]
\caption{Security requirements discussion.}{
\begin{tabular}  { | p {2 cm} |   p {15 cm}|}
\hline
\textbf{Requirement }     & \textbf{Employed method} \\\hline
Confidentiality & Encryption (symmetric or asymmetric) is used for all transactions (Sections \ref{sec:sub-overlay} and \ref{sec-smarthome}).  \\\hline
Integrity  & Each transaction includes a hash of all other fields contained in the transaction  (Sections \ref{sec:sub-overlay} and \ref{sec-smarthome}). \\\hline
Availability   & 1) LBM processes all incoming transactions and controls access to all smart home devices, thus protecting them from malicious requests, 2) An OBM sends a transaction to its cluster members only if a key contained in the transaction matches one of the entries in its keylist (Section \ref{sec:sub-overlay}).  This ensures  that the cluster members only receive transactions from authorized nodes. \\\hline
Authentication  &  In the smart home,  the LBM creates a shared key between two communicating devices which is also used for authentication (Section \ref{sec-smarthome}).  In the overlay, each node should have a stored genesis transaction in the BC to be authenticated. As transactions are chained to the genesis transaction, a node is authenticated when it has the private  key corresponding to the output PK of a transaction stored in the BC (Section \ref{sec:sub-overlay}). The cloud storage uses standard authentication protocols.   \\\hline
Non-repudiation  & Overlay transactions are signed by the transaction generator to achieve non-repudiation. Additionally, all overlay transactions are stored in the public BC, so neither requester nor requestee can deny their complicity in a transaction (Section \ref{sec:sub-overlay}).  \\\hline
\end{tabular}}\label{tab:security-req}
\end{table*}
In Table \ref{tab:attack evaluation} we summarize 12 specific security attacks to which IoT networks or BCs are particularly vulnerable  and outline how LSB protects against them.  In Table  \ref{tab:attack evaluation} we also analyse how resilient LSB is against each attack and the likelihood of the attack to happen  based on  European Telecommunications Standards Institute (ETSI) \cite{etsi2011102} risk analysis criteria. These criteria evaluate each attack  based on the following five metrics: i) \textit{time:} the cumulative time for an attacker to first detect a vulnerability, and subsequently plan and launch a successful attack, ii) \textit{expertise:} the generic expertise that the attacker must possess about the underlying principles in order to orchestrate the attack, iii) \textit{knowledge:}  specific information that is available about the target system, e.g., security configuration, iv) \textit{opportunity:} the duration and nature (e.g., continuous or intermittent) of access to the system needed for launching the attack,  v) \textit{equipment:} software and/or hardware necessary for conducting the attack. LSB exhibits beyond high resistance  to seven attacks and high resistance to three attacks. This suggests that LSB is highly secure. \par 
\begin{table*}
\caption{Studying attacks on BlockChain. }{
\begin{tabular}  { | p {1.5 cm} | p {4 cm}|  p {8 cm}| p {1 cm}| p {1 cm}|}
\hline
 \textbf{Attack }    & \textbf{Definition}  & \textbf{Defence} & \textbf{Resistant to attack} & \textbf{Attack likelihood} \\\hline
Appending  attack & Attacker compromises an OBM and generates blocks with fake transactions to create false reputation  & An OBM can detect a fake block during the verification step (see Section \ref{sec:sub:verification}) when it verifies the output and owner of the ledger   & Beyond high & Unlikely\\\hline
 Denial of Service (DOS) attack  &  Attacker floods  an overlay node (target) with a large number of transactions to overwhelm the node such that it cannot devote any resources to process genuine transactions from other nodes. &  i) OBMs would not send a transaction to their cluster members unless they find  a match with an entity in their keylist, ii) Each overlay node has a threshold for the maximum rate of transactions received from the overlay. If the threshold is exceeded, the keylist is updated to prevent nodes from sending transactions to the target node. Details  are discussed in Section \ref{sec:sub-overlay}.  &High & Unlikely  \\\hline
Distributed DOS (DDOS) attack & This is a distributed version of the above attack, where multiple overlay nodes or smart home devices are compromised by the attacker.  &  1) Infecting devices and overlay nodes is difficult due to usage of OBM keylists and the fact that devices are not directly accessible (see Sections \ref{sec:sub-overlay},\ref{sec-smarthome}); 2) In a smart home the LBM authorizes all transactions prior to sending to the overlay (see Section \ref{sec-smarthome}); 3) The overlay transactions  are not valid within the smart home, as different encryption methods are employed in the overlay and the smart home  (see Sections  \ref{sec:sub-overlay},\ref{sec-smarthome}); 4) In the smart home, a device can only communicate with another device if a shared key has been established between them by the LBM (see Section \ref{sec-smarthome}). Finally, the methods that prevent DOS attacks in the above row are also useful for mitigating DDOS. & Beyond high & Unlikely \\\hline
Device injection attack & Attacker introduces fake devices to the smart home to gain access to private information within the home. & The injected device is isolated  as local communications require that the LBM has set up a shared key, which requires approval from the home owner as discussed in Section \ref{sec-smarthome}. & Moderate & Possible  \\\hline
Linking attack & Attacker (that can be a SP or cloud storage) links  multiple data  in the cloud or transactions in the BC with the same ID  to find the real world identity of an anonymous node.  &  Overlay nodes  use a unique PK for each transaction in the overlay. Each device is authenticated with the cloud using separate accounts. This prevents the attacker from linking the data of multiple devices of the same user.  & Beyond high & Unlikely \\\hline
Dropping attack & The OBM drops  transactions to or from its cluster members to isolate them  from the overlay &  A cluster member can change the OBM it is associated with if it observes that its transactions are not being processed. & High & Unlikely\\\hline
Modification attack & Malicious cloud storage changes or removes stored data  & The store transaction includes the hash of stored data,  as shown in Fig \ref{fig:transactions-flowchart}, that serves as evidence of when the data was stored or last modified; however, data can not be recovered. & Beyond high & Unlikely \\\hline
False reputation & An overlay node increases its reputation by increasing  output[0] in each of its transactions by more than one & OBMs detects false increase during the transaction verification discussed in Algorithm \ref{algo-verification}. & Moderate  & Possible \\\hline
Public BC modification & Attacker advertises a false ledger of blocks   and makes it  as the longest ledger. Thus, all nodes accept the attacker ledger as the true ledger.  & The consensus algorithm limits the number of blocks that an OBM can generate within a time interval. This will limit the number of malicious blocks that an OBM can append, and thus prevent the attacker from generating a longer ledger than the true ledger.  (see Section \ref{sec:consensus}).  & Beyond high & Unlikely \\\hline
Breaking the time interval  & Malicious OBM generates more than one block in each consensus-period & Other OBMs would detect this as they would receive more than the permitted number of blocks in a consensus-period (see Section \ref{sec:consensus}). Consequently, the trust rating for this OBM would be decreased and it would be isolated from the rest of the overlay.  & Beyond high & Unlikely \\\hline
Consensus-period attack & The attacker(s) sends false requests to update the consensus-period & For a request to be considered valid, it must be signed by at least half the number of OBMs. The likelihood of this is very low. & High & Unlikely \\\hline
 51\% attack & The attacker controls more than 51\% of OBMs and tries to compromise the consensus algorithm by generating fake blocks or more than the permitted number of blocks & The attack can be detected during block verification (see Algorithm \ref{algo-verification}) or by other OBMs based on consensus algorithm (see Section \ref{sec:consensus}). & Beyond high & Unlikely \\\hline
\end{tabular}}\label{tab:attack evaluation}
\end{table*}%

\textit{\textbf{Privacy:}} LSB uses anonymity and user control to protect the privacy of users in the   overlay, smart home and the cloud storage. Using changeable PKs as the identity of overlay nodes introduces similar level of anonymity and privacy  as is experienced in  other BC-based systems (e.g. Bitcoin). In certain IoT applications, the two end points that are communicating may need to know the real identity of  each other. For example, a home insurance company needs to know the real identity of the owner of the smart home that it is insuring. In these instances,  the corresponding transaction generator uses  a unique  PK to  communicate with each overlay node.  Stored transactions in the public BC are encrypted using the requestee PK to protect the privacy of the overlay nodes    against attackers who attempt to read the data in metadata field of a multisig transaction (see Fig \ref{fig:overlay-trans-struc}). \par 
In the smart home, the LBM enforces home owner policies to ensure  his control over exchanged data, thus protects his privacy. \par
The cloud storage is able to use the data of different devices of an overlay node to find his real word identity. To protect against this de-anonymization of devices, the overlay node uses different credits to store the data of each of its devices. This prevents the cloud to identify different devices of the same overlay node.   \par 

\textbf{Fault tolerance: } Fault tolerance is a measure of how resilient an architecture is to node failures. It is evident from Section \ref{sec:def&cons} that LBM and OBMs implement various key functions and the failure of these nodes could thus potentially impact the normal operation of LSB. The failure of an LBM would disconnect the corresponding smart home and the associated devices from the overlay. The smart devices would be able to share data locally but would not be able to store data in the cloud storage or communicate with other overlay devices. \par 
In case   an OBM leaves the overlay, the cluster members associated with this OBM would not receive any service. However, they can readily select a new OBM to associate with. The OBM(s) departure may also affect the overlay throughput as   there are fewer OBMs to generate blocks. However, the DTM mechanism outlined in \ref{sub-sec:dis-throughput} can handle this situation. The departure of multiple OBMs may also impact security due to the corresponding actions of the distributed trust mechanism. Recall from Section \ref{sec:sub:verification} that as OBMs garner trust in one another, fewer transactions within a block need to be verified. Thus, when multiple OBMs leave, the probability of detecting a fake transaction in a new block decreases as fewer OBMs remain in the network to validate the new blocks. We will further elaborate on the minimum number of OBMs required to participate in the BC to prevent attacks in section \ref{eval:4-trust}.\par 
 \subsection{Performance evaluation}
 In this section, we present extensive evaluations of various performance aspects of LSB. We first explored the possibility of using open source BC instantiations such as Ethereum. However, these platforms are particularly suited for developing applications on top of the underlying BC substrate. LSB has significant differences in its fundamental operations in comparison to these BC instantiations. As a result, we were unable to use these platforms for our evaluations and thus chose to use simulations.   We evaluate the smart home tier and the overlay separately since these tiers operate independently, particularly when pertaining to performance. Thus, we abstract over the details of one tier when studying the performance of the other tier. We use the following two simulators: \par 
 \textit{Cooja}: We use Cooja \cite{coojaref} to study the performance of the smart home tier. Cooja  is well-suited for evaluating low resource devices and benefits the availability of implementation of various IoT-aware protocols. \par 
 \textit{NS3}: We use NS3 \cite{NS3} to evaluate the  overlay performance as it has been widely used for analysing peer-to-peer networks.\par 
For the NS3 simulations, we consider a network consisting of 50 overlay nodes. We assume the $T\_max$ to be 10. We assume five requesters generate four transactions per second. The above settings are referred to as the default configuration and are used in the simulations unless explicitly noted otherwise.  \par 
In the rest of this section, we first evaluate the POW processing time in Section \ref{eval:1pow}. The smart home tier performance is evaluated in Section \ref{eval:2-smart home}. Next, we evaluate the delay  which an overlay node experiences while requesting  smart home data in Section \ref{eval:3accessing-home}.  Distributed trust and its effects on the overlay security and performance  are studied in Section \ref{eval:4-trust}.  Finally, we evaluate DTM in Section \ref{eval:5-throughput}.
\subsubsection{POW processing time} \label{eval:1pow}
In this part of evaluation,  we  aim to evaluate the time consumed by an off-the-shelf device to solve the POW, one of the widely used consensus algorithms in BC-based systems \cite{gervais2016security}. We do so to highlight the ineffectiveness of using classic BC and PoW in the IoT context. Each block in the Bitcoin BC has a nonce attached to them. The miner is required to search for the correct nonce such that the block as a whole satisfies a certain arbitrary condition. Specifically, it is required that the SHA-256 hash of the block have a certain number of leading zeros. The only way to find the correct nonce is by brute force. The number of leading zeros controls the difficulty of solving the POW. The longer the length of this sequence, the more resources and processing time required to solve the puzzle.  We implemented POW using C++ on a MacBook Pro ( 2.7 Ghz Intel Core i5 processor, 8 GB memory, and Intel Iris Graphics 6100 graphic card)  to study the processing time of solving the puzzle with two difficulties. Typical  IoT devices are significantly more resource-constrained than a laptop so the results obtained are conservative upper bounds  that one can expect with IoT devices. Solving PoW with 6 leading zeros takes 2.3 seconds. Increasing the length of zeros to 7, increases the processing time to 29.22 minutes. Currently, Bitcoin is using blocks with 17 zeros, which would take exponentially longer to solve on a standard laptop. The results presented herein confirm that solving the POW  as used in Bitcoin incurs significant delays on laptop class devices, thus validating our design choice of eliminating PoW in LSB.
\subsubsection{Smart home performance}\label{eval:2-smart home}
We conduct simulations using Cooja to evaluate the energy consumption and time overhead  of  the LBM. This is because the LBM   is the most resource consuming device  in the smart home since it  handles all transactions and performs many hashing and encryption (both symmetric and aymmetric) operations.  In contrast, the IoT devices have to perform very simple tasks of which the most computationally intensive is symmetric encryption. It has been shown \cite{mukhopadhyay2014internet} that most IoT devices have sufficient capabilities to perform this task.  A complete evaluation of the smart home tier is provided in our previous work \cite{dorri2017BC}. In this paper, we present a summary of these results.  To compare the overhead of LSB, we simulate another method that has the same transaction flow as LSB,  but does not use  encryption, hashing, and  local IL. We refer to this  as the "baseline". We use IPv6 over Low Power Wireless Personal Area Networks (6LoWPAN)  as the underlying communication protocol in our simulation, since it is well-suited to the resource constraints for a smart home setting. We simulate three z1 mote sensors (that mimic smart home devices) which send data directly to the LBM (also simulated as a z1 mote) every 10 seconds. Each simulation lasts for 3 minutes and the results  are averaged over this duration.  A cloud storage is  directly connected to the LBM for storing data. To provide a comprehensive evaluation, we simulate store and access transactions. For the store transaction, we simulated two different and realistic traffic flow patterns:  \par 

\textit{Periodic}: In this setting, devices periodically  store  data on the cloud storage (similar to a smart thermostat storing temperature readings periodically in the cloud).  \par 
\textit{Query-based}: Herein, the devices  store data when they received a query  from the user, e.g. a home owner queries a connected security camera to check whether anyone approached the door. \par 

We evaluate the following metrics: \par 
\begin{itemize}
\item \textit{Time overhead:} Refers to the processing time for each transaction in the LBM and is measured from when a transaction is received at the LBM  until the appropriate response is sent to the requester. \par 
\item \textit{Energy consumption:}  Refers to the energy consumed by the LBM for processing transactions.  \par 
\end{itemize}
\par
\begin{figure} [b]  	
\includegraphics[ width=8cm ,height=8cm, keepaspectratio]{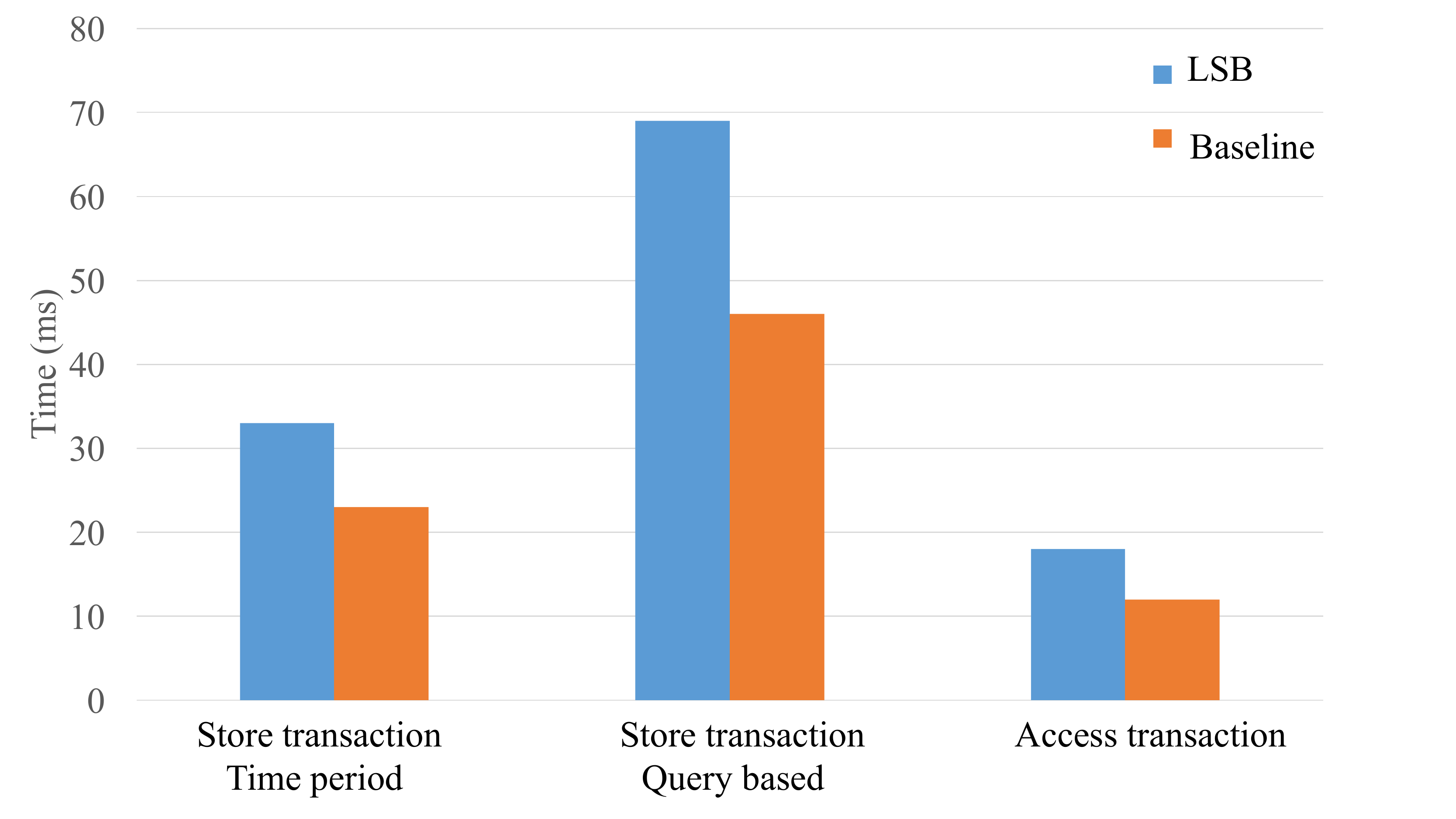}
\caption{Evaluation of time overhead in the LBM.}
\label{fig:per-home-time}
\end{figure}

\textit{Time overhead:}   Fig \ref{fig:per-home-time} shows the results for the time overhead. LSB consumes more time to process packets compared to the baseline  which can be attributed to the additional encryption and hashing operations. In the worst case for the query-based store transaction, the additional overhead introduced by LSB is 20ms, which is still small in absolute terms.\par
 
\begin{figure} [h]  	
\includegraphics[ width=8cm ,height=8cm, keepaspectratio]{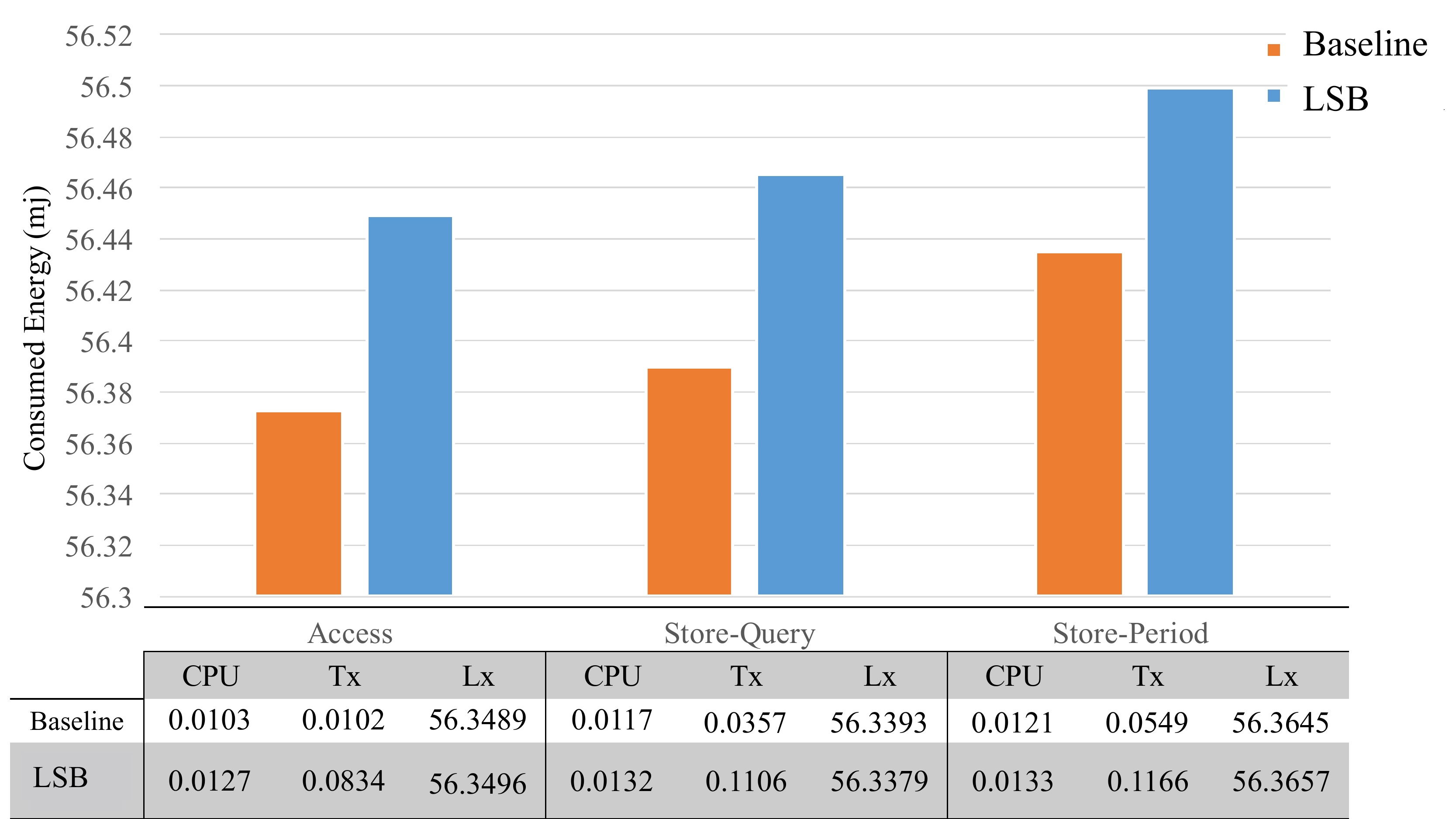}
\caption{Evaluation of energy consumption.}
\label{fig:energy-cons}
\end{figure}

\textit{ Energy consumption:}  Fig \ref{fig:energy-cons} outlines the energy consumption results. As is evident, LSB increases the energy consumption by 0.07 (mj). The table at the bottom of Fig \ref{fig:energy-cons} outlines the energy consumption for the 3 core tasks performed by the LBM, namely: CPU, transmission (Tx), and listening (Lx).
The energy consumption by CPU increases by approximately  0.002(mj) in LSB due to encryption and hashing. LSB results in longer packets (due to encryption and hashing), which doubles the transmission energy consumption as compared with the baseline.  It should be noted that we have assumed a 100\% radio duty cycle in our evaluations (i.e. the radio is always on). If the radio is switched off intermittently to conserve energy, then the relative listening overhead incurred by LSB  would be higher. However, even assuming a very aggressive duty cycle of 1\%, the relative increase in listening energy would still only be about 60\%.  \par 
\subsubsection{Accessing a smart home device from the overlay}\label{eval:3accessing-home}
In this section, we evaluate end-to-end delay experienced by an overlay node for accessing or monitoring a smart home device (e.g. when a home owner is remotely connected to the overlay and wishes to monitor his security camera at home).  Delay is measured from the time since the request is generated till the response is received. We conduct simulations using NS3 with the default configuration with 13 overlay nodes acting as OBMs. We compare LSB with a baseline method which is consistent to current smart home offerings on the market, where the requester directly communicates with the LBM without the need for any of the transaction processing that is part of LSB. The delay incurred using the baseline method is 17.62 ms. On the contrary, with LSB, the delay increases to 48.74ms. The higher delay can be attributed to the fact that the transaction has to be broadcast to other OBMs for verification. Each OBM incurs a delay of 0.006ms for processing the transaction (the precise steps are outlined in Section \ref{sec:sub-overlay}).  However, this delay is relatively insignificant.   \par

\begin{figure} [h]  	
\includegraphics[ width=8cm ,height=8cm, keepaspectratio]{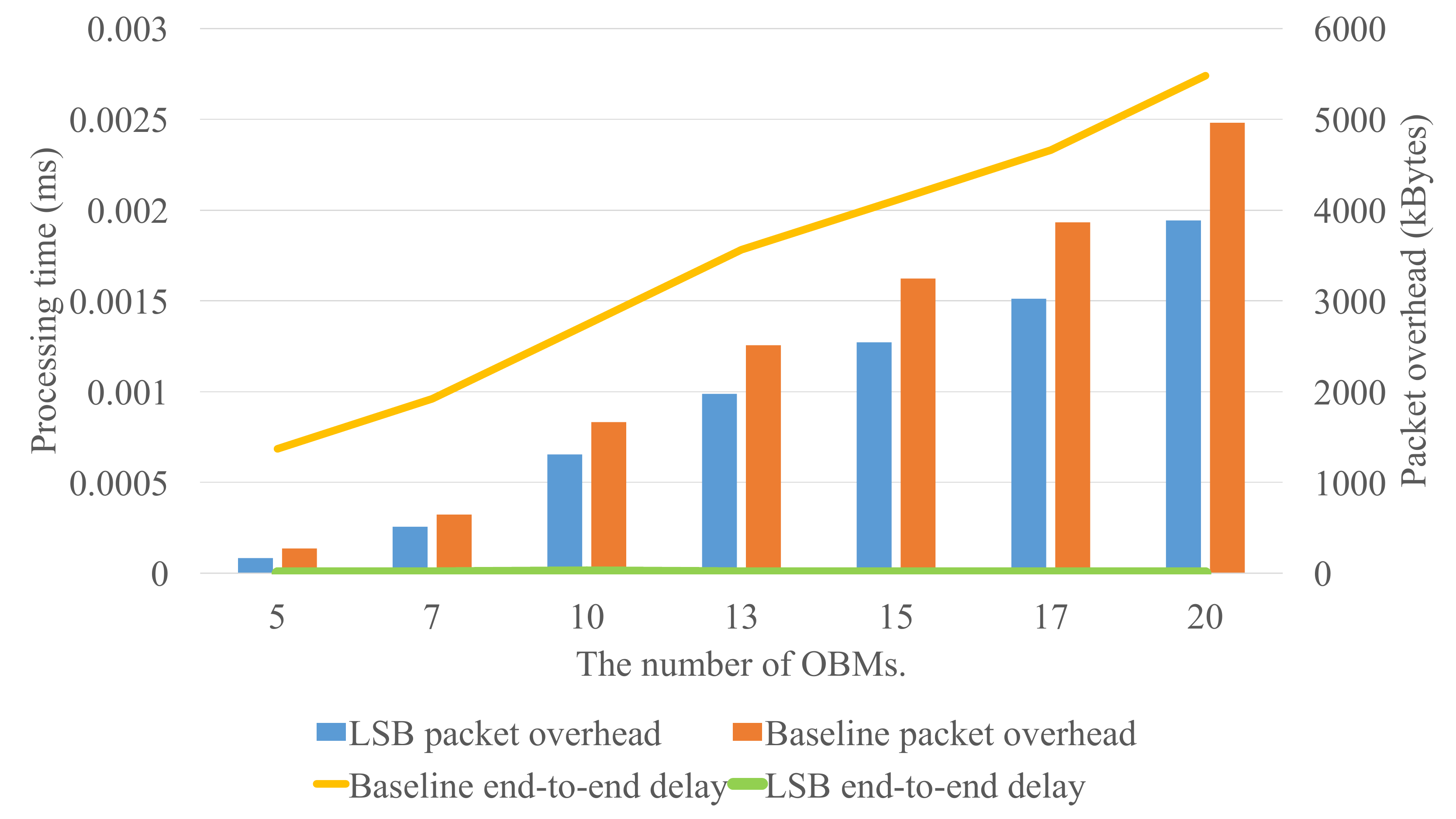}
\caption{Assessing the impact of separating the data and transaction flows.}
\label{fig:per-seperation-allandone}
\end{figure}

As was noted in Sections \ref{sec:sub-overlay} and \ref{sec:transaction-flow}, LSB separates the data flow from the transaction flow. While transactions are broadcast amongst the OBMs in the overlay, the data packets are forwarded towards the destination along optimal paths as determined by a routing protocol such as OSPF. To quantify the benefits of this design decision, we compare LSB with a baseline method wherein both the transactions and data packets are broadcast in the overlay network. We use the default configuration and assume that one requester sends four access transactions  per second to a requestee. We consider the following two performance metrics which are best at capturing the impact of the separation between the transaction and data flows: (i) end-to-end delay - similar to above (ii) packet overheads - this captures the total number of packets transmitted by OBMs  for delivering the data packets to the requester. Since, the size of the data and transaction packets are different, we measure the latter as the cumulative sum of all packet sizes in KBytes. Since, these two metrics are affected by the number of OBMs in the network, we vary the number of overlay nodes that act as OBMs from 5 to 20. The results are presented in Fig. \ref{fig:per-seperation-allandone}.   LSB incurs lower packet overhead and end-to-end delay compared to the baseline since, in the latter the data packets are broadcast among all OBMs as compared to the former where the data packets are routed along optimal paths. Observe that, for the baseline, both metrics grow linearly as the number of OBMs increase. The amount of broadcast traffic generated is directly proportional to the number of OBMs which explains the linear increase in the packet overhead. Since the data packets are now broadcast, the delay incurred in receiving the data at the requester also increases linearly. In contrast, with LBS only the packet overhead increases with the number of OBMs. Since the data packets are routed directly to the requester, the end-to-end delay is not affected by the number of OBMs. \par 
These results demonstrate the efficacy of keeping the data and transaction flows independent of each other. \par

\subsubsection{Evaluation of the Distributed Trust Algorithm}\label{eval:4-trust}
\begin{figure}[h]   	
\includegraphics[ width=9cm ,height=9cm, keepaspectratio]{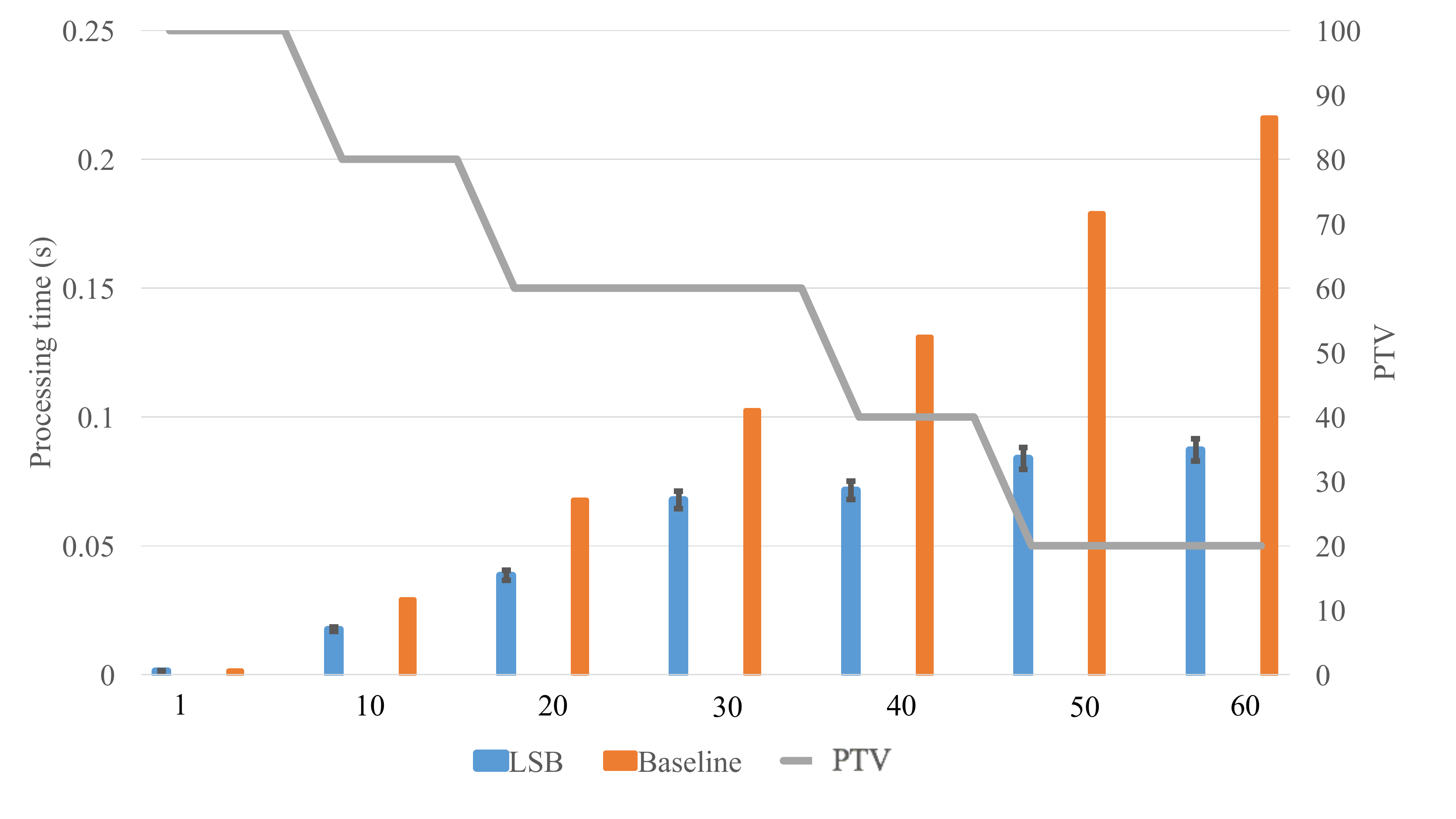}
\caption{The average processing time on OBMs to validate new blocks.}
\label{fig:per-overlay-process}
\end{figure}

Recall that in the classical BC, all transactions within a new block must be verified by an overlay node. In contrast, LSB uses a distributed trust algorithm wherein the number of transactions that must be verified decreases gradually as OBMs build up trust in each other (see Section \ref{sec:sub:verification}). In this experiment, we compare the processing time for validating a new block in LSB with a baseline strategy that is similar to classical BCs. We use the the default network configuration and the trust table shown in Fig \ref{fig:trust-overlay-pic}. The simulation lasts for 180 seconds and the results, shown in Fig \ref{fig:per-overlay-process},  are the average of 10 runs. The standard deviation is also shown, except for the baseline where results are deterministic.   We measure the time taken by each OBM to validate a new block and plot the average in Fig \ref{fig:per-overlay-process} (shown on the left vertical axis). Note that, we disregard all other tasks (e.g. checking key lists, generating new blocks, etc) other than validation of new blocks in this evaluation as the former are not affected by the trust algorithm.  Fig \ref{fig:per-overlay-process}  plots the processing time as a function of the number of blocks successfully verified (and thus appended to the BC) as the simulation progresses. The percentage of transactions that need to be verified (PTV) is shown on the right vertical axis. As can be inferred from Fig \ref{fig:per-overlay-process}, at start up, the processing time is the same for both methods since the OBMs have yet to garner trust in each other. However, as time progresses and more blocks are generated and verified, the OBMs build up  direct  trust in each other. Consequently, only a fraction of the total transactions in a new block need to verified in LSB, which reduces the processing time as compared to the baseline, wherein all transactions within the block are verified. Moreover, as the number of blocks verified increases, progressively less transactions need to be verified (also shown in Fig \ref{fig:per-overlay-process}) as the trust in other OBMs continues to increase.  Once 50 blocks are generated, the trust  among OBMs reaches the highest level (see Fig \ref{fig:trust-overlay-pic}). From here on, the number of transactions that need to be verified remain fixed and so does the processing time. At steady state (i.e., when the network has been running for a substantial period of time), LSB achieves over 50\% savings in processing time compared to the baseline. \par

\begin{figure}[h]   	
\includegraphics[ width=9cm ,height=9cm, keepaspectratio]{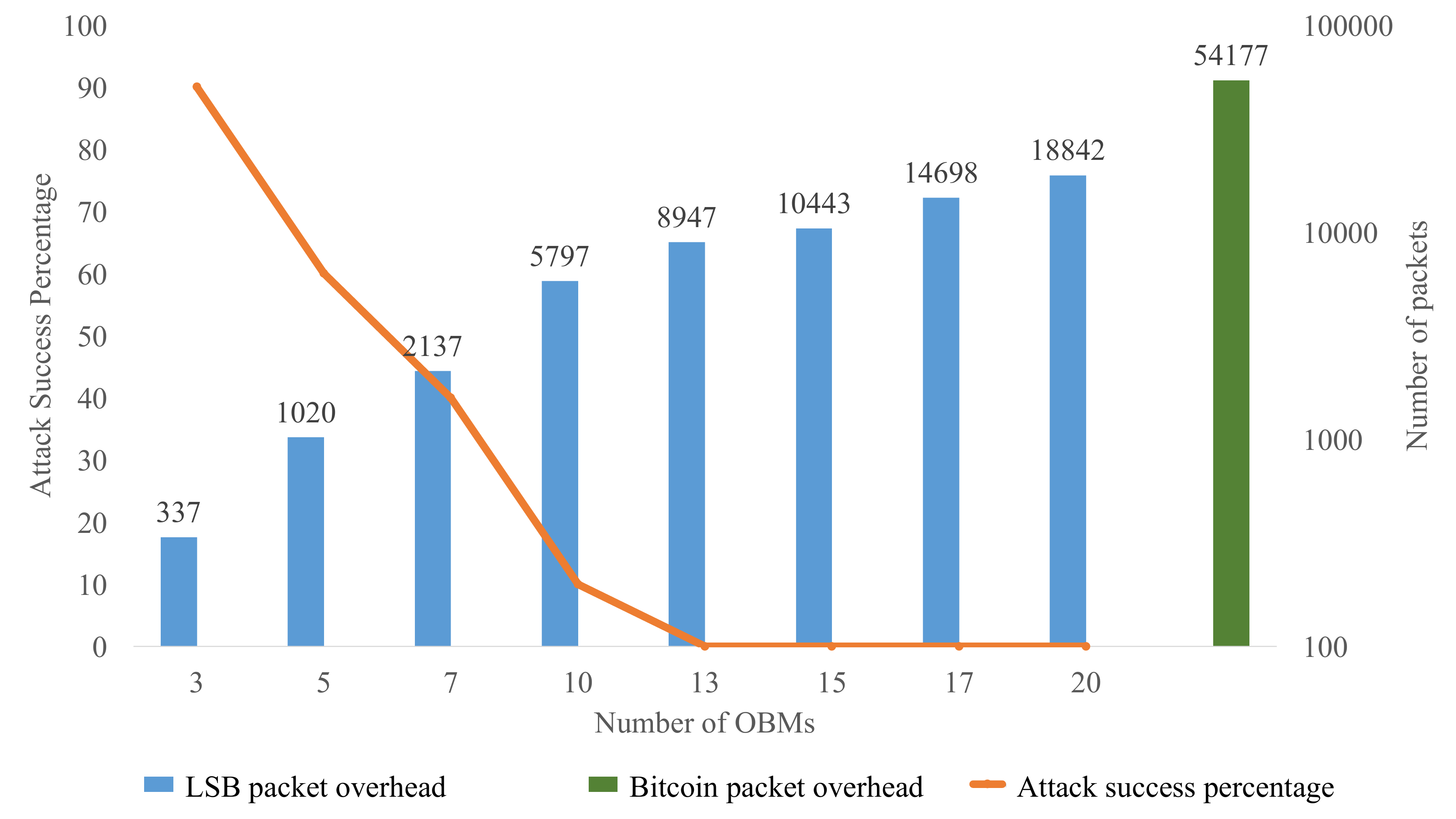}
\caption{Evaluating the impact of the number of OBMs  on security and  packet overhead.}
\label{fig:per-overlay-security-trust}
\end{figure}

In LSB, since only a fraction of transactions within a block are verified, there is a chance that a fake transaction created by malicious node may not be verified and thus appended to the BC (referred to as appending attack in Section \ref{sec:sub:security-analysis}). In the following, we evaluate the success percentage of such an attack. Intuitively, the more OBMs in the network, the lower the likelihood of a successful attack, since the chance that the fake transaction will be picked for verification increases. However, the packet overhead also increases proportionally with the number of OBMs due to the increase in the broadcast traffic. To study this trade-off, we consider the default network configuration and vary the number of overlay nodes acting as OBMs from 3 to 20. The evaluation metrics are the attack success rate and the cumulative packet overhead.  To simulate the attack, we consider the worst-case scenario, where a highly trusted OBM, which has generated more than 50 blocks and has thus accrued a high level of trust, creates a new block containing one fake transaction. We use the trust table shown in Fig \ref{fig:trust-overlay-pic}. We run the simulation 10 times and attack success is the percentage of the number of runs that the fake block is  not detected by any of the honest OBMs (this applies to all evaluations that consider security attacks  in the rest of the paper). We compare the packet overheads incurred in LSB with a baseline wherein the overlay network is structured similarly to Bitcoin. Recall that in Bitcoin all overlay nodes (50 in our case) manage the BC distributedly unlike LSB where BC management is limited to  selected overlay nodes, i.e., OBMs. Note that, the baseline would always accurately detect the attack, since all transactions in a block are verified. The results are shown in Fig \ref{fig:per-overlay-security-trust}.  Observe that, as the number of OBMs increases, the likelihood of a successful attack reduces substantially. As expected, the packet overhead increases linearly with the number of OBMs. With 13 OBMs, all attacks are successfully detected. However, the corresponding packet overhead (8497) is significantly lower than that incurred in the baseline (54177).\par 
\begin{figure}[h]   	
\includegraphics[ width=9cm ,height=9cm, keepaspectratio]{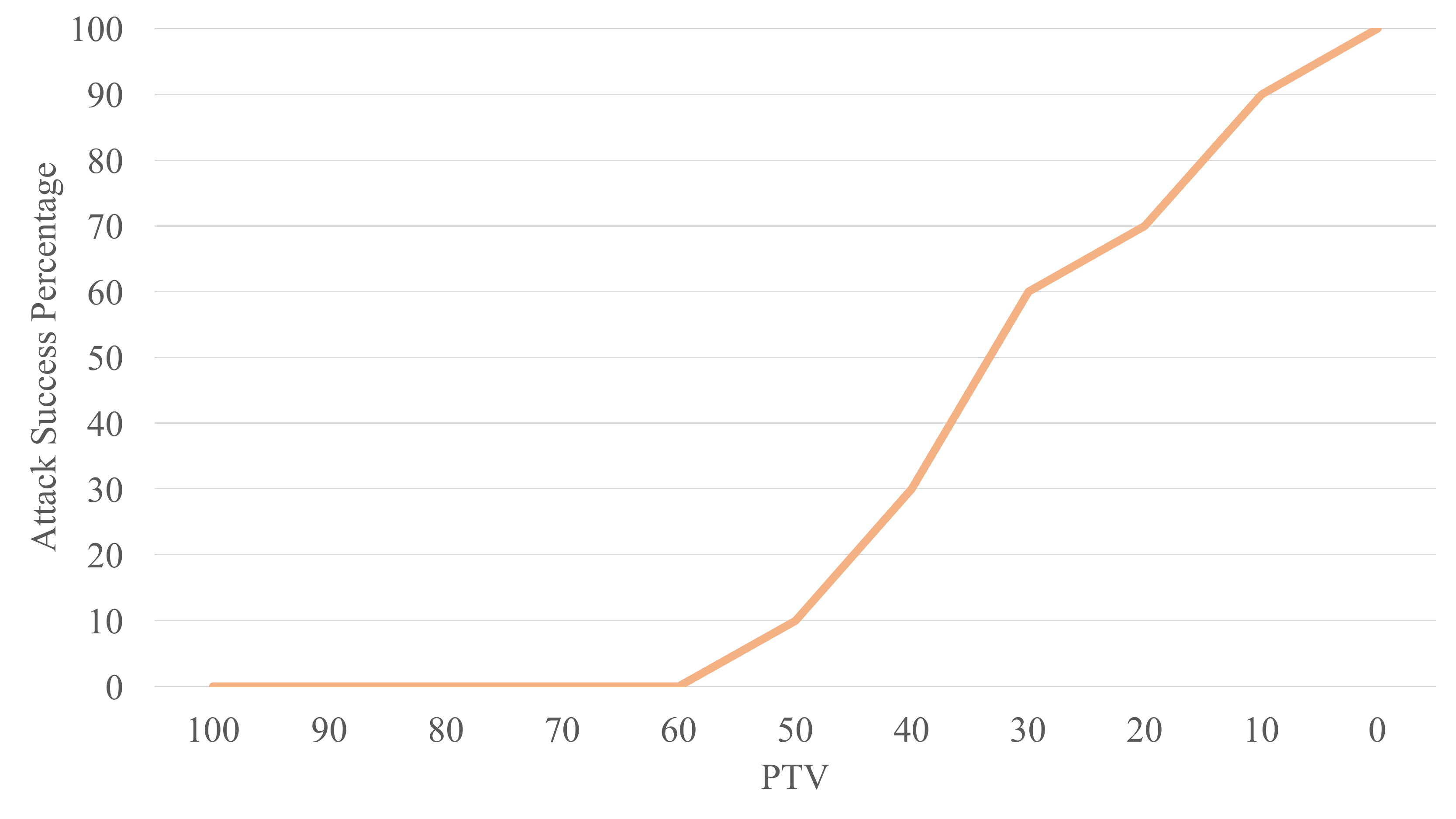}
\caption{Evaluating the impact of PTV on the ability to detect appending attacks.}
\label{fig:per-overlay-processing-trust}
\end{figure}

The attack success percentage is directly impacted by the PTV. To study this impact, we evaluate the attack success percentage  for different PTVs in a network with default configuration with five overlay nodes acting as OBMs. The reason for choosing five OBMs is to show the effect of PTV on the attack success. As is evident from Fig \ref{fig:per-overlay-security-trust}, the presence of a greater number of OBMs improves the security considerably and these effects are not as evident. The results are illustrated in Fig \ref{fig:per-overlay-processing-trust}. When PTV equals to 100, OBMs  verify all transactions in a block, leading to zero attack success percentage (i.e., the attack is always detected). As shown in Fig \ref{fig:per-overlay-processing-trust}, as PTV  decreases, the attack success percentage  increases. For the network configuration used in this simulation, the lowest value of PTV that can guarantee security is 60.  We have repeated the same simulation for different number of OBMs to determine the smallest value of PTV for which an attack can always be detected. Table \ref{tab:max-trust-level} shows the results of the simulation and can be used as a guideline to configure the trust table (e.g. Fig \ref{fig:trust-overlay-pic}). A PTV that is lower than the values in Table \ref{tab:max-trust-level}  will make the network vulnerable to appending attacks. On the other hand, a larger value increases the processing time for new blocks (as more transactions need to be verified) and the packet overhead in the network.  \par 

\begin{table}[h]
\caption{Minimum PTV for detecting appending attacks as a function of the number of OBMs.}{
\begin{tabular}  {  | p {2.68 cm}| p {0.3 cm} |   p {0.3 cm}|   p {0.3 cm} |   p {0.3 cm} |   p {0.3 cm}|   p {0.3 cm}|   p {0.3 cm}|   p {0.3 cm} |}
\hline
\textbf{Number of OBMs  }     & 3 & 5 & 7& 10 & 13 & 15 & 17 & 20 \\\hline
\textbf{Minimum PTV } & 80  &60 & 60 & 40&20 & 20& 20 & 10 \\\hline
\end{tabular}}\label{tab:max-trust-level}
\end{table}

\subsubsection{DTM performance analysis}\label{eval:5-throughput}
\begin{figure}[h]   	
\includegraphics[ width=9cm ,height=9cm, keepaspectratio]{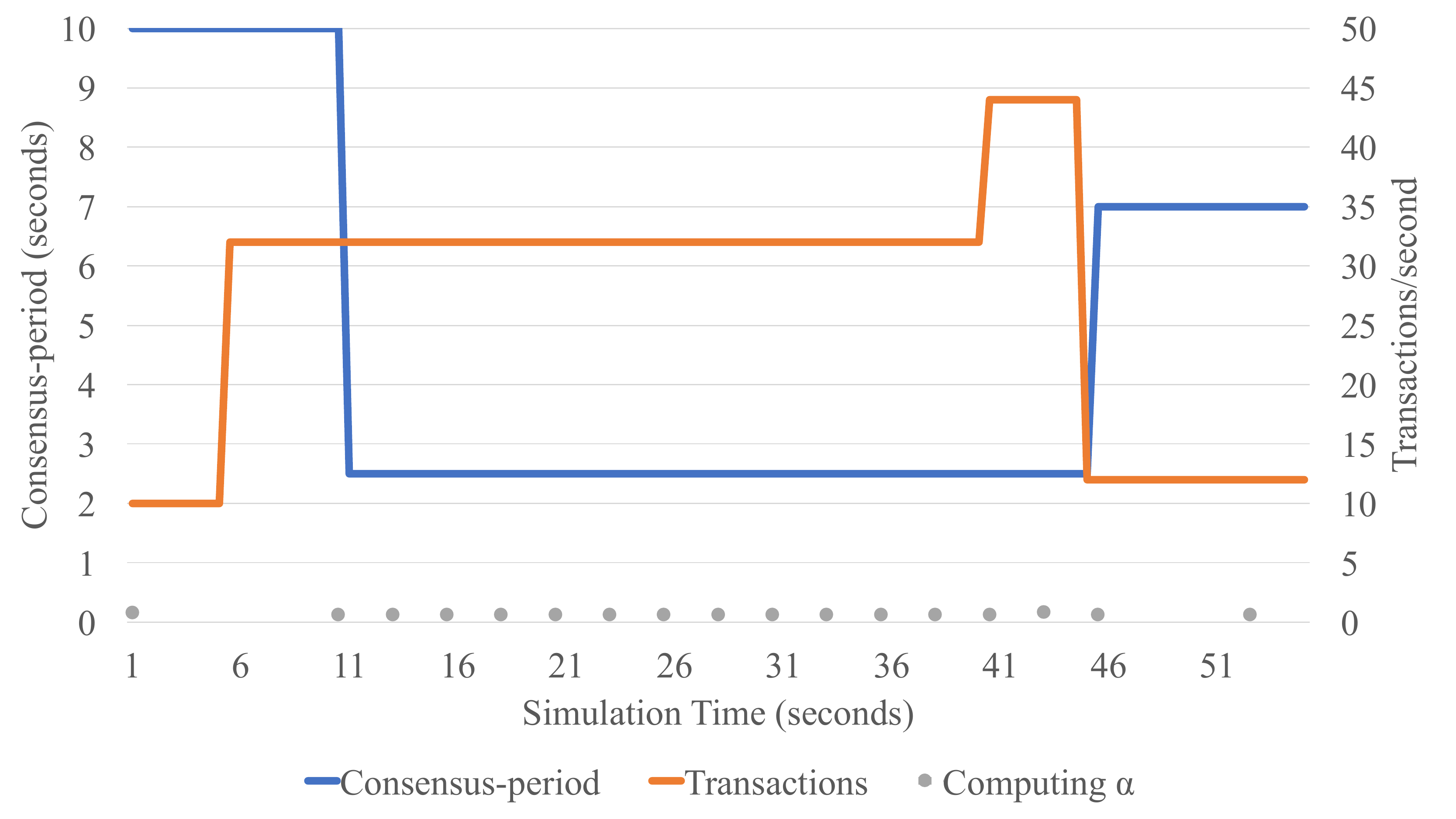}
\caption{Evaluation of DTM in the overlay. }
\label{fig:per-overlay-time-period-increase}
\end{figure}

The DTM mechanism proposed in Section \ref{sub-sec:dis-throughput}  aims to dynamically adjust the network utilization based on the total load, i.e., the number of generated transactions. To  illustrate the performance of DTM, we simulate a network with the default configuration with 13 overlay nodes acting as OBMs.     As classical BCs have fixed throughput (e.g., the Bitcoin BC has a fixed throughput of 7 transactions per second) there is no baseline  that we can use for comparison. Initially, a total of 10 cumulative transactions are generated in the overlay network per second. We simulate situations where the network demand fluctuates. The number of transactions per second increases to 32 for the entire time period from 5 seconds to 40 seconds. At 40 seconds, the load increases further to 44 transactions per second until 45 seconds when the load drops to 12 transactions per second. The changes in the network load are illustrated in Fig \ref{fig:per-overlay-time-period-increase}. Recall that DTM  computes the network utilization, $ \alpha $, at the end of each consensus-period as the ratio between the  number of transactions generated and the number of transactions added to the overlay BC since the last computation of $\alpha$. Time intervals  when $ \alpha $ is computed by OBMs is shown using gray dots in the figure.  The consensus period is initially set to 10 seconds, which is the default value. We assume that $\alpha_{min}$ and $\alpha_{max}$  are set to 0.25 and 1, respectively. \par 
At the end of the first consensus-period (i.e. 10 seconds), $\alpha $ is computed  to be 2.4, which is greater than $\alpha_{max}$ (1). This is because of the sharp increase in the network load at 5 seconds. To reduce the network utilization, DTM reduces the consensus-period to the newly computed value of 2.5 seconds (see lines 2-5 Algorithm \ref{algo-DTM}) using Equation \ref{formula}, where $\alpha $ is set to 0.62, which is the mid point of  $\alpha_{min}$ and $\alpha_{max}$.  The consensus-period is also illustrated in Fig \ref{fig:per-overlay-time-period-increase}. Subsequently, since the network load remains stable until 40 seconds, the consensus period also remains unchanged. At this time, the  network load increases further. Thus, at the end of the next consensus-period (43 seconds),   $\alpha $ is computed to be 0.84. As the computed value is still less than $\alpha_{max}$, no further action is required.  This highlights the effectiveness of choosing the mid-point of $\alpha_{min}$ and $\alpha_{max}$ for recomputing  $\alpha $.  The value of   $ \alpha $ drops to 0.2 at 48 seconds resulting from the sharp decrease in the number of transactions at  45 seconds. Since this value is less than $\alpha_{min}$, DTM increases the consensus-period (see lines 7-9 Algorithm \ref{algo-DTM}) to a new value, which is computed as 7 seconds.

\section{Discussions }\label{discussion}
\subsection{Proof of activities} 
Recall that (Section \ref{sec:def&cons}), all transactions pertaining to devices within a smart home are recorded and stored in the local IL as a device-specific ledger.  This information can be very useful for auditing purposes, i.e., to precisely track who and when had access to a specific sensor and its data. In order to ensure that the local IL has not been tampered with, we recommend that a hash of it is periodically stored in the public BC using a single signature transaction (discussed in Section \ref{sec:sub-overlay}). In the following, we outline a use case demonstrating the auditing functionality. One can readily envision other scenarios. \par 
Consider a situation where a law enforcement agency requires proof that a particular individual, say Alice, was at home during a particular time period, e.g., last night. Assume that Alice's home is equipped with a smart doorbell camera and motion sensors and that all data is stored in the local storage. The camera would have captured and stored footage of Alice arriving home on that day and subsequently leaving her home on the next day. Additionally, the motion sensors would have also logged Alice's presence at home. Alice can thus present this data and the associated transactions in the IL to the investigators to confirm her presence at home on that particular night. Additionally, the investigators can verify the hash of Alice's IL stored in the public BC to ensure that the data presented has not been modified.

\subsection{Shared overlay}
In instances where an individual is responsible for several homes (e.g., when he owns multiple homes or is supporting family members such as elderly parents), managing each residence separately can multiply the effort. LSB offers the option of creating a shared overlay to simplify the management of multiple smart homes. The manager chooses one LBM in one home as the Shared BM (ShBM) for all homes. Other homes in the shared overlay use a Virtual Private Network (VPN) connection (between the home Internet gateway and the ShBM) to connect and share data with the ShBM securely. The ShBM performs the same functions as the LBM but for all devices in all the homes that are part of the shared overlay.
\subsection{OBM reward}
In classical BCs, e.g. Bitcoin or Ethereum, nodes that generate new blocks are offered a monetary reward in the form of coins as a form of compensation for expending their resources to solve the computationally intensive puzzle associated with block creation. This fee is paid as the transaction fee by the users. However, there is now a growing consensus that for more effective BC the transaction fee should be removed  \cite{zdnet}.  LSB employs a lightweight consensus algorithm and thus we do away with explicit rewards and the transaction fee.  Instead, an OBM on generating a valid block gains reputation with other OBMs (see Section \ref{sec:def&cons}), which could be construed as an implicit reward. \par 
Another way to incentivize OBMs could be to allow them to place advertisements in the blocks that they append to the BC. An explicit field within the block header could be reserved for this purpose. The advertisement is also included in the  block hash which prevents other OBMs from changing the advertisements.  This may be particularly attractive to service and cloud storage providers. \par 
 \section{Related Works} \label{relatedwork}
In this section, we provide a literature review on IoT security and privacy and BC-based systems.  \par 
\textit{\textbf{IoT security: }} Authors in \cite{6990267} proposed an end-to-end  host identity protocol to secure  IoT. The proposed method reduces the header size of the 6LowPAN and Host Identity protocol (HIP)  from 40 bytes to a maximum of 25 bytes by eliminating unnecessary header fields and thus reduces network overhead.  The authors also proposed a lightweight key distribution method for distributing keys between low resource IoT devices and users. A high resource available device is placed in the wireless range of the low resource devices to perform resource consuming tasks on behalf of the low resource devices.   Although their approach is computationally lightweight for their considered particular application, removing the 6LowPAN and HIP header fields  leads to reduced functionality. Moreover, the scalability of this approach is limited due to the fact that the high resource device must be within wireless range of all IoT devices. \par 
The authors in \cite{6258209} proposed a new authentication and access control method to make  IoT secure against unauthorized users and access. The proposed method relies on two authentication authorities namely: i) Registration Authority (RA), and ii) Home Registration Authority (HRA). The RA is  designed to facilitate the authentication process for  devices. All devices are registered with the RA.  Similarly, the HRA facilitates the authentication process for the users. When a user wishes to access data from a particular device, the request is first sent to the RA. The RA checks the authenticity of the user with the HRA. Assuming the user is authenticated, the RA generates a shared key for communication between the user and the device.  Security analysis shows that the proposed method is secure against the man-in-the-middle attack.  However, the need for each device to have a RA and correspondingly each user to have a HRA could be a bottleneck for scalability. In LSB, we have rather proposed a tiered structure where a single public BC is managed distributedly by the overlay nodes and the devices within each smart home are managed independently by a home-specific LBM.  Our approach scales better while also achieving protection against a broader range of attacks. \par 
\textit{\textbf{BC applications}}: The notion of a BC was first introduced in the landmark paper \cite{nakamoto2008bitcoin}  on Bitcoin by Satoshi Nakamoto. Bitcon aims to  do away with centralized authorities for money exchange while offering a high level of security and privacy to the users.    In 2013 a new BC platform, called Ethereum, was introduced  \cite{wood2014ethereum}. Ethereum users are able to generate smart contracts with a small fee but with high security and privacy. Several applications have been proposed in recent years that make use of the Ethereum BC including BC in agriculture \cite{FullProfile}, crowd funding \cite{Wei-fund}, and micro blogging \cite{eth-twitter}.\par 
Numerous other applications of BCs have been proposed recently.  Authors in \cite{aitzhan2016security} proposed a novel application of BC in energy trading. Using their proposed framework, energy producers can negotiate the selling price with their customers and also facilitate a smart contract to make a sale. A Distribution System Operator (DSO) ensures that the trade is secure and prevents the possibility for either a producer or customer to not follow through with their part of the contract. A lock key is used to prevent an energy producer from double spending (i.e. selling the energy to more than one customer). Security analysis shows that the framework is secure to a broad range of attacks. However, the architecture suffers from low scalability as a result of  broadcasting  all transactions and blocks to the whole network. In LSB, we overcome this challenge by limiting the number of nodes who manage the BC. \par 
The authors in \cite{hashemi2016world} proposed a BC-based multi-tier architecture to share data from IoT devices with organizations and people. The proposed architecture has three main components namely: data management protocol, data store system, and message service. The data management protocol provides a framework for data owner, requester, or data source to communicate with each other. The messaging system is used to increase the network scalability based on a publish/subscribe model.  Finally, the data store system uses a BC for storing data privately. As in our work, they do not rely on POW given the associated overheads.  In contrast to this work, we do not use the BC for storing user data as it will consume large bandwidth to store data in the distributed BC. Instead, we store hash of the data in the cloud in the public BC. \par 
Recently Intel has designed a new consensus algorithm for BC known as Proof of Elapsed Time (POET) which is integrated with Hyperledger \cite{baliga2017understanding}.  POET is a leader election algorithm which is intended to run in a Trusted Execution Environment (TEE) in Intel CPUs. Before a node can store a block in the BC, it must wait for a random time which is selected from a trusted enclave. A TimeChecker function verifies the choice of the random time. The block can only be appended to the BC after this time period. The consensus protocol (see Section \ref{sec:consensus}) used in LSB is conceptually similar to POET. However, LSB does not rely on a particular hardware platform and is thus more generalized.  \par 
The authors in  \cite{popov2016tangle} proposed a new ledger based  cryptocurrency called IoTA.  By eliminating the notion of blocks and mining, IoTA ensures that the transactions are free and verification is fast. The key innovation behind IoTA is the "tangle", which is essentially a directed acyclic graph (DAG). Before a user can send a transaction, he has to verify two randomly chosen transactions generated by other users. As the number of nodes increase, the transactions generated also increase but so do the number of transactions that are verified. LSB shares some similarities with IoTA  such as zero transaction fees and both realize a self-scaling network. However, LSB employs a BC unlike the DAG employed by IoTA. LSB thus benefits from the inherent benefits of a BC such as the auditability offered by an immutable ledger.   

\section{Conclusion} \label{conclusion}
 In this paper, we argued that although BlockChain (BC) is an effective technology for providing security and privacy in IoT, its application in the IoT context presents several significant challenges including: complexity, bandwidth and latency overheads and scalability. To address these challenges, we proposed a Lightweight Scalable BC (LSB) for IoT. LSB has an IoT  friendly consensus algorithm  that eliminates the need for solving any puzzle prior to appending a block to the BC. LSB incorporates a distributed trust method whereby the processing time for validating new blocks by the OBMs gradually decreases as they build up trust in each other. A distributed throughput management strategy adjusts certain system parameters to ensure that the network utilization is within a prescribed operating range.  Security analysis demonstrates that LSB is highly secure against a broad range of attacks.  In the instance when key nodes fail, LSB operation exhibits graceful degradation, thus making it highly fault tolerant.  Simulation results show that the proposed architecture  decreases bandwidth and processing time  compared to the classical BCs. Additionally, the smart home owners  receive services with no additional delay for local transactions (i.e. smart home communications) and with a small imperceptible delay for the overlay transactions compared to the state-of-the-art.   Generally,  LSB brings a high level of security and privacy for IoT users while enforcing a marginal overhead. As the future direction of this work, one may consider relaxing some of the assumptions made in LSB, e.g. considering cases when the local block manager might be compromised.\par 
In our future work, we plan to develop a prototype implementation of LSB to understand its performance in real-world settings. We will also explore the suitability of LSB in other application domains such as smart grids and vehicular networks.  

\section{Acknowledgements}
The authors are thankful to Mr. Raghunath Ganta from Tata Consultancy Services for his insightful feedback.

\bibliographystyle{IEEEtran}
\bibliography{IEEEabrv,bare_jrnl_compsoc} 

\begin{thebibliography}{10}
\providecommand{\url}[1]{#1}
\csname url@samestyle\endcsname
\providecommand{\newblock}{\relax}
\providecommand{\bibinfo}[2]{#2}
\providecommand{\BIBentrySTDinterwordspacing}{\spaceskip=0pt\relax}
\providecommand{\BIBentryALTinterwordstretchfactor}{4}
\providecommand{\BIBentryALTinterwordspacing}{\spaceskip=\fontdimen2\font plus
\BIBentryALTinterwordstretchfactor\fontdimen3\font minus
  \fontdimen4\font\relax}
\providecommand{\BIBforeignlanguage}[2]{{%
\expandafter\ifx\csname l@#1\endcsname\relax
\typeout{** WARNING: IEEEtran.bst: No hyphenation pattern has been}%
\typeout{** loaded for the language `#1'. Using the pattern for}%
\typeout{** the default language instead.}%
\else
\language=\csname l@#1\endcsname
\fi
#2}}
\providecommand{\BIBdecl}{\relax}
\BIBdecl

\bibitem{kosba2016hawk}
A.~Kosba, A.~Miller, E.~Shi, Z.~Wen, and C.~Papamanthou, ``Hawk: The blockchain
  model of cryptography and privacy-preserving smart contracts,'' in
  \emph{Security and Privacy (SP), 2016 IEEE Symposium on}.\hskip 1em plus
  0.5em minus 0.4em\relax IEEE, 2016, pp. 839--858.

\bibitem{nakamoto2008bitcoin}
S.~Nakamoto, ``Bitcoin: A peer-to-peer electronic cash system,'' 2008.

\bibitem{wood2014ethereum}
G.~Wood, ``Ethereum: A secure decentralised generalised transaction ledger,''
  \emph{Ethereum Project Yellow Paper}, vol. 151, 2014.

\bibitem{yue2016healthcare}
X.~Yue, H.~Wang, D.~Jin, M.~Li, and W.~Jiang, ``Healthcare data gateways: Found
  healthcare intelligence on blockchain with novel privacy risk control,''
  \emph{Journal of medical systems}, vol.~40, no.~10, p. 218, 2016.

\bibitem{abramaowicz2016cryptocurrency}
M.~Abramaowicz, ``Cryptocurrency-based law,'' \emph{Ariz. L. Rev.}, vol.~58, p.
  359, 2016.

\bibitem{BCInIndustry}
B.~I. O. T. S. . B. I. W. B. C.~B. Used,
  \url{https://www.cbinsights.com/blog/industries-disrupted-blockchain},
  [Online; accessed 19-April-2017].

\bibitem{vukolic2015quest}
M.~Vukoli{\'c}, ``The quest for scalable blockchain fabric: Proof-of-work vs.
  bft replication,'' in \emph{International Workshop on Open Problems in
  Network Security}.\hskip 1em plus 0.5em minus 0.4em\relax Springer, 2015, pp.
  112--125.

\bibitem{Altcoin}
Altcoin, \url{http://altcoins.com}, [Online; accessed July-2017].

\bibitem{ferrer2016blockchain}
E.~C. Ferrer, ``The blockchain: a new framework for robotic swarm systems,''
  \emph{arXiv preprint arXiv:1608.00695}, 2016.

\bibitem{brambilla2016using}
G.~Brambilla, M.~Amoretti, and F.~Zanichelli, ``Using block chain for
  peer-to-peer proof-of-location,'' \emph{preprint arXiv:1607.00174}, 2016.

\bibitem{zhang2014iot}
Z.-K. Zhang, M.~C.~Y. Cho, C.-W. Wang, C.-W. Hsu, C.-K. Chen, and S.~Shieh,
  ``Iot security: ongoing challenges and research opportunities,'' in
  \emph{Service-Oriented Computing and Applications (SOCA), 2014 IEEE 7th
  International Conference on}.\hskip 1em plus 0.5em minus 0.4em\relax IEEE,
  2014, pp. 230--234.

\bibitem{de2014openpds}
Y.-A. de~Montjoye, E.~Shmueli, S.~S. Wang, and A.~S. Pentland, ``openpds:
  Protecting the privacy of metadata through safeanswers,'' \emph{PloS one},
  vol.~9, no.~7, p. e98790, 2014.

\bibitem{dorri2017towardsBC}
A.~Dorri, S.~S. Kanhere, and R.~Jurdak, ``Towards an optimized blockchain for
  iot,'' in \emph{IEEE/ACM International conference on Internet-of-Things
  Design and Implementation (IoTDI)}, 2017.

\bibitem{dorri2017BC}
A.~Dorri, S.~S. Kanhere, R.~Jurdak, and P.~Gauravaram, ``Blockchain for iot
  security and privacy: The case study of a smart home,'' in \emph{IEEE Percom
  workshop on security privacy and trust in the Internet of Things}.\hskip 1em
  plus 0.5em minus 0.4em\relax IEEE, 2017.

\bibitem{kousaridas2015systas}
A.~Kousaridas, S.~Falangitis, P.~Magdalinos, N.~Alonistioti, and M.~Dillinger,
  ``Systas: Density-based algorithm for clusters discovery in wireless
  networks,'' in \emph{Personal, Indoor, and Mobile Radio Communications
  (PIMRC), 2015 IEEE 26th Annual International Symposium on}.\hskip 1em plus
  0.5em minus 0.4em\relax IEEE, 2015, pp. 2126--2131.

\bibitem{stoica2001chord}
I.~Stoica, R.~Morris, D.~Karger, M.~F. Kaashoek, and H.~Balakrishnan, ``Chord:
  A scalable peer-to-peer lookup service for internet applications,'' \emph{ACM
  SIGCOMM Computer Communication Review}, vol.~31, no.~4, pp. 149--160, 2001.

\bibitem{OSPFref}
R.~. Open Shortest Path~First(OSPF),
  \url{https://www.ietf.org/rfc/rfc2328.txt}, [Online; accessed
  19-November-2016].

\bibitem{lu2008framework}
K.~Lu, Y.~Qian, M.~Guizani, and H.-H. Chen, ``A framework for a distributed key
  management scheme in heterogeneous wireless sensor networks,'' \emph{IEEE
  Transactions on Wireless Communications}, vol.~7, no.~2, 2008.

\bibitem{ongaro2014search}
D.~Ongaro and J.~K. Ousterhout, ``In search of an understandable consensus
  algorithm.'' in \emph{USENIX Annual Technical Conference}, 2014, pp.
  305--319.

\bibitem{Bogdanov2011}
A.~Bogdanov, M.~Kne{\v{z}}evi{\'{c}}, G.~Leander, D.~Toz, K.~Var{\i}c{\i}, and
  I.~Verbauwhede, \emph{spongent: A Lightweight Hash Function}.\hskip 1em plus
  0.5em minus 0.4em\relax Berlin, Heidelberg: Springer Berlin Heidelberg, 2011,
  pp. 312--325.

\bibitem{fsecuresense}
F.-S. sense, \url{www.sense.f-secure.com}, [Online; accessed Nov-2016].

\bibitem{delfs2002introduction}
H.~Delfs, H.~Knebl, and H.~Knebl, \emph{Introduction to cryptography}.\hskip
  1em plus 0.5em minus 0.4em\relax Springer, 2002, vol.~2.

\bibitem{cooper2008internet}
D.~Cooper, ``Internet x. 509 public key infrastructure certificate and
  certificate revocation list (crl) profile,'' 2008.

\bibitem{tschorsch2015bitcoin}
F.~Tschorsch and B.~Scheuermann, ``Bitcoin and beyond: A technical survey on
  decentralized digital currencies,'' \emph{IEEE Communications Surveys \&
  Tutorials}, vol.~18, no.~3, pp. 2084--2123, 2015.

\bibitem{etsi2011102}
T.~ETSI, ``102 165-1 v4. 2.3 (2011-03),'' \emph{Technical Specification
  Telecommunications and Internet converged Services and Protocols for Advanced
  Networking (TISPAN)}, 2011.

\bibitem{coojaref}
Cooja, \url{http://anrg.usc.edu/contiki/index.php/CoojaSimulator}, [Online;
  accessed 19-November-2016].

\bibitem{NS3}
NS3, \url{https://www.nsnam.org}, [Online; accessed July-2017].

\bibitem{gervais2016security}
A.~Gervais, G.~O. Karame, K.~W{\"u}st, V.~Glykantzis, H.~Ritzdorf, and
  S.~Capkun, ``On the security and performance of proof of work blockchains,''
  in \emph{Proceedings of the 2016 ACM SIGSAC Conference on Computer and
  Communications Security}.\hskip 1em plus 0.5em minus 0.4em\relax ACM, 2016,
  pp. 3--16.

\bibitem{mukhopadhyay2014internet}
S.~C. Mukhopadhyay and N.~Suryadevara, ``Internet of things: Challenges and
  opportunities,'' in \emph{Internet of Things}.\hskip 1em plus 0.5em minus
  0.4em\relax Springer, 2014, pp. 1--17.

\bibitem{zdnet}
ZDNet, ``Zdnet,''
  \url{http://www.zdnet.com/article/a-better-blockchain-bitcoin-for-nothing-and-transactions-for-free/},
  2017.

\bibitem{6990267}
S.~Sahraoui and A.~Bilami, ``Compressed and distributed host identity protocol
  for end-to-end security in the iot,'' in \emph{2014 International Conference
  on Next Generation Networks and Services (NGNS)}, May 2014, pp. 295--301.

\bibitem{6258209}
J.~Liu, Y.~Xiao, and C.~L.~P. Chen, ``Authentication and access control in the
  internet of things,'' in \emph{32nd International Conference on Distributed
  Computing Systems Workshops}, 2012, pp. 588--592.

\bibitem{FullProfile}
FullProfile, \url{www.fullprofile.com.au}, [Online; accessed July-2017].

\bibitem{Wei-fund}
Wei-fund, \url{www.weifund.io}, [Online; accessed July-2017].

\bibitem{eth-twitter}
eth twitter, \url{www.github.com/yep/eth-tweet}, [Online; accessed July-2017].

\bibitem{aitzhan2016security}
N.~Z. Aitzhan and D.~Svetinovic, ``Security and privacy in decentralized energy
  trading through multi-signatures, blockchain and anonymous messaging
  streams,'' \emph{IEEE Transactions on Dependable and Secure Computing}, 2016.

\bibitem{hashemi2016world}
S.~H. Hashemi, F.~Faghri, P.~Rausch, and R.~H. Campbell, ``World of empowered
  iot users,'' in \emph{2016 IEEE First International Conference on
  Internet-of-Things Design and Implementation (IoTDI)}.\hskip 1em plus 0.5em
  minus 0.4em\relax IEEE, 2016, pp. 13--24.

\bibitem{baliga2017understanding}
A.~Baliga, ``Understanding blockchain consensus models,'' Tech. rep.,
  Persistent Systems Ltd, Tech. Rep., 2017.

\bibitem{popov2016tangle}
S.~Popov, ``The tangle,'' \emph{cit. on}, p. 131, 2016.

\end{thebibliography}
\begin{IEEEbiography} [{\includegraphics[width=1in,height=1.25in,clip,keepaspectratio]{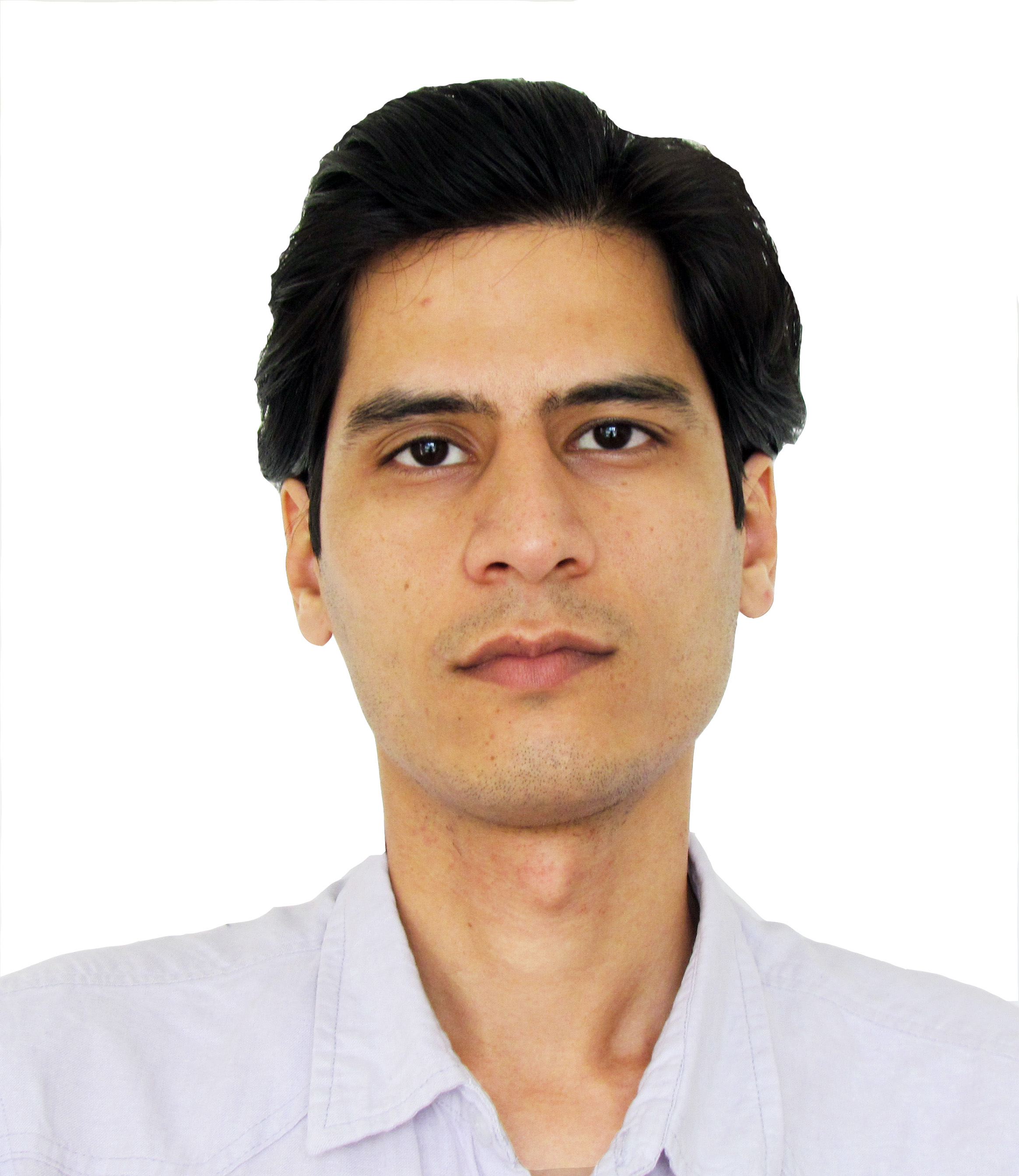}}]{Ali Dorri}
 received his bachelor degree in Computer Engineering from Bojnourd University, IRAN, 2012. He then commenced his master degree in Computer Engineering in Islamic Azad University of Mashhad, IRAN, working on Mobile Ad hoc Networks and security issues rising from this sort of network. He now is a Ph.D. candidate in UNSW, Sydney. His current research interest covers security and privacy concerns in the context of Internet of Things (IoT), Wireless Sensor Network (WSN) and Vehicular Ad hoc Network (VANET). Moreover, he is working on blockchain and its applications on IoT.
\end{IEEEbiography}
\vspace{-10 mm}
\begin{IEEEbiography} [{\includegraphics[width=1in,height=1.25in,clip,keepaspectratio]{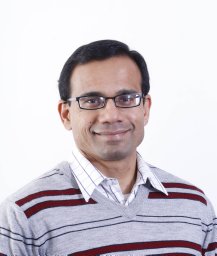}}]{Salil S. Kanhere }
 received his M.S. and Ph.D. degrees, both in Electrical Engineering from Drexel University, Philadelphia. He is currently an Associate Professor in the School of Computer Science and Engineering at UNSW, Sydney, Australia. His current research interests include Internet of Things, pervasive computing, crowdsourcing,  privacy and security. He has published over 150 peer-reviewed articles and delivered over 20 tutorials and keynote talks on these research topics. He is a contributing research staff at Data61, CSIRO. Salil regularly serves on the organizing committee of a number of IEEE and ACM international conferences. He currently serves as the Area Editor for Pervasive and Mobile Computing and Computer Communications. Salil is a Senior Member of both the IEEE and the ACM. He is a recipient of the Humboldt Research Fellowship in 2014. 
\end{IEEEbiography}

\begin{IEEEbiography} [{\includegraphics[width=1in,height=1.25in,clip,keepaspectratio]{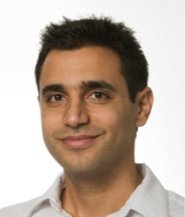}}]{Raja Jurdak}
is a Senior Principal Research Scientist at CSIRO, where he leads the Distributed Sensing Systems Group. He has a PhD in Information and Computer Science at University of California, Irvine. His current research interests focus on energy-efficiency, mobility, and security in networks. He has over 120 peer-reviewed journal and conference publications, as well as a book published by Springer in 2007 titled Wireless Ad Hoc and Sensor Networks: A Cross-Layer Design Per- spective. He regularly serves on the organizing and technical program committees of international conferences (DCOSS, RTSS, Sensapp, Percom, EWSN, ICDCS). Raja is an Honorary Professor at University of Queensland, and an Adjunct Professor at Macquarie University and James Cook University. He is a Senior Member of the IEEE. ​
\end{IEEEbiography}

\begin{IEEEbiography} [{\includegraphics[width=1in,height=1.25in,clip,keepaspectratio]{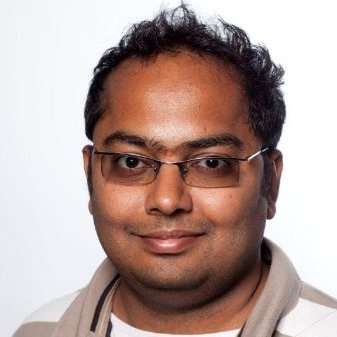}}]{Praveen Gauravaram}
 is a consultant and scientist in cyber security at Tata Consultancy Services (TCS), Australia 
and an Adjunct Senior Lecturer at University of New South Wales, Australia. Praveen's focus is on embedding 
innovation \& creativity into TCS's customer deliverables and offerings. Praveen has a PhD in Cryptology from Queensland University of Technology, Brisbane, Australia. Praveen has held scientific positions in India, Europe and Australia and is a recipient of research grants and awards whist his postdoctoral fellowship at Technical University of Denmark. Praveen’s significant scientific achievements include co-design of Grostl cryptographic hash function 
a finalist in the SHA3 cryptographic hash competition conducted by US National Institute of Standards and Technology and security evaluation of standard cryptographic designs. Praveen has led a team of researchers at TCS Innovation Labs Hyderabad on analysing performance aspects of fully homomorphic encryption for secure cloud computing. To date, Praveen has published more than forty research papers in several conferences, journals and research consortiums. 

\end{IEEEbiography}

\end{document}